\documentclass[prl,twocolumn,superscriptaddress,showpacs,amsmath,amssymb]{revtex4-1}

\usepackage[dvipsnames]{xcolor}
\usepackage{graphicx}
\usepackage{bm}
\usepackage[colorlinks=true,allcolors=blue]{hyperref}
\usepackage{comment}

\definecolor{TTH-color2}{named}{Maroon}
\definecolor{RO-color2}{named}{Green}
\definecolor{MS-color2}{named}{TealBlue}

\newcommand{\timeorder}{\mathbb{T}}
\newcommand{\contourorder}{\mathbb{T}}

\begin{document}

\title{Spin currents driven by the Higgs mode  in magnetic superconductors }

\author{Mikhail A. Silaev}
\affiliation{Department of Physics and Nanoscience Center, University of Jyvaskyla, P.O. Box 35 (YFL), FI-40014 University of Jyvaskyla, Finland}
\affiliation{Moscow Institute of Physics and Technology, Dolgoprudny, 141700 Russia}

\author{Risto Ojaj\"arvi}
\affiliation{Department of Physics and Nanoscience Center, University of Jyvaskyla, P.O. Box 35 (YFL), FI-40014 University of Jyvaskyla, Finland}

\author{Tero T. Heikkil\"a}
\affiliation{Department of Physics and Nanoscience Center, University of Jyvaskyla, P.O. Box 35 (YFL), FI-40014 University of Jyvaskyla, Finland}

\date{\today}

\begin{abstract}
Higgs mode in superconducting materials describes slowly-decaying oscillations of the order parameter amplitude.
We demonstrate that in magnetic superconductors with built-in spin-splitting field Higgs mode is strongly coupled to the spin degrees of freedom allowing for the generation of time-dependent spin currents. 
Converting such spin currents
to electric signals by spin-filtering elements provides a tool for the second-harmonic generation and the  electrical detection of the Higgs mode generated by the external irradiation. The non-adiabatic spin torques generated by these spin currents allow for the 
magnetic detection of the Higgs mode by measuring the precession of
magnetic moment in the adjacent ferromagnet. We discuss also the
reciprocal effect which is the generation of the Higgs mode by the 
magnetic precession. 
Coupling the collective modes in superconductors to light and  magnetic dynamics opens the new direction of superconducting optospintronics. 
\end{abstract}

\maketitle


Oscillations of the 
order parameter amplitude  in condensed matter systems are often called Higgs modes \cite{Varma2002,Podolsky2011,Pashkin2014,Volovik2014,Varma2015} by analogy with the Higgs boson in particle physics \cite{Higgs1964}.  These collective excitations are generic for  ordered states like antiferromagnets, charge density waves\cite{RevModPhys.60.1129}, superfluids
 \cite{PhysRevLett.30.829,PhysRevLett.30.541, Zavjalov2016}, cold atomic gases \cite{PhysRevLett.106.205303,Endres2012} and superconductors \cite{Volkov1973,Sooryakumar1980,littlewood1982amplitude, Barankov2004,grasset2018higgs,Matsunda2013,Matsunaga1145, Sherman2015,Podolsky2011,tsuji2015theory, PhysRevB.96.020505, PhysRevLett.120.117001, 1809.10335, 1809.06989,silaev2019nonlinear}. 

 In general the experimental observation of Higgs modes is quite challenging. 
 They have been  
 observed by Raman scattering in superconductors with charge density wave order \cite{Sooryakumar1980,PhysRevB.26.4883, grasset2018higgs, PhysRevLett.122.127001} and by the nuclear magnetic resonance in superfluid $^3$He \cite{PhysRevLett.30.829,PhysRevLett.30.541, Zavjalov2016}.
 In usual superconductors without extra broken symmetries probing the collective modes has become possible only recently owing to the development of 
 low-temperature THz spectroscopy
 \cite{PhysRevLett.107.177007, 
 beck2013transient, Matsunaga2012,Matsunda2013,Matsunaga1145,Giorgianni2019}.
 Measuring nonlinear optical responses in THz domain  allows 
 for probing collective modes in several superconducting compounds 
 \cite{Matsunda2013,Matsunaga1145,PhysRevB.96.020505,Giorgianni2019}.
 For example in NbN the Higgs mode frequency given by twice the order parameter amplitude $2\Delta$ \cite{Volkov1973,  
 Kulik1981,Barankov2004,Barankov2006} lies in the THz range and therefore can be observed using optical probes, including the pump-probe technique
 \cite{Matsunda2013}  and the resonant third-harmonic generation in the transmitted signal
 \cite{Matsunaga1145}. 

 In this paper we show that Higgs modes can be observed
 through the generated spin currents using purely electrical probes in 
 the wide class of magnetic superconductors. 
  Unusual spin transport properties of such systems have attracted intense attention recently \cite{RevModPhys.90.041001, Beckmann2016},
  stimulating both
  experimental \cite{Huebler2012,Wolf2014,
  1906.09079,DeSimoni2018,
  PhysRevMaterials.1.054402,PhysRevLett.116.097001,
PhysRevB.90.144509,PhysRevB.87.024517,Quay2013,Quay2016} and 
  theoretical efforts
  \cite{ Silaev2015,Bobkova2015,Krishtop2015,PhysRevB.93.014512,PhysRevB.98.024516,RevModPhys.90.041001,PhysRevB.98.020501,PhysRevB.93.024513}. 
   The underlying physical mechanism behind the suggested electrical measurement of the Higgs mode  is rooted in the strong coupling between the superconducting order parameter dynamics and electron spins.
   The possibility to transmit spin signals by the order parameter excitations has been elucidated using the example of mobile  topological defects - Abrikosov vortices \cite{PhysRevLett.121.187203, Vargunin2019}. 
 Here we demonstrate that time-dependent spin currents can be generated by the collective amplitude modes in superconductors. 
  %
  \begin{figure}
\includegraphics[width=.45\columnwidth]{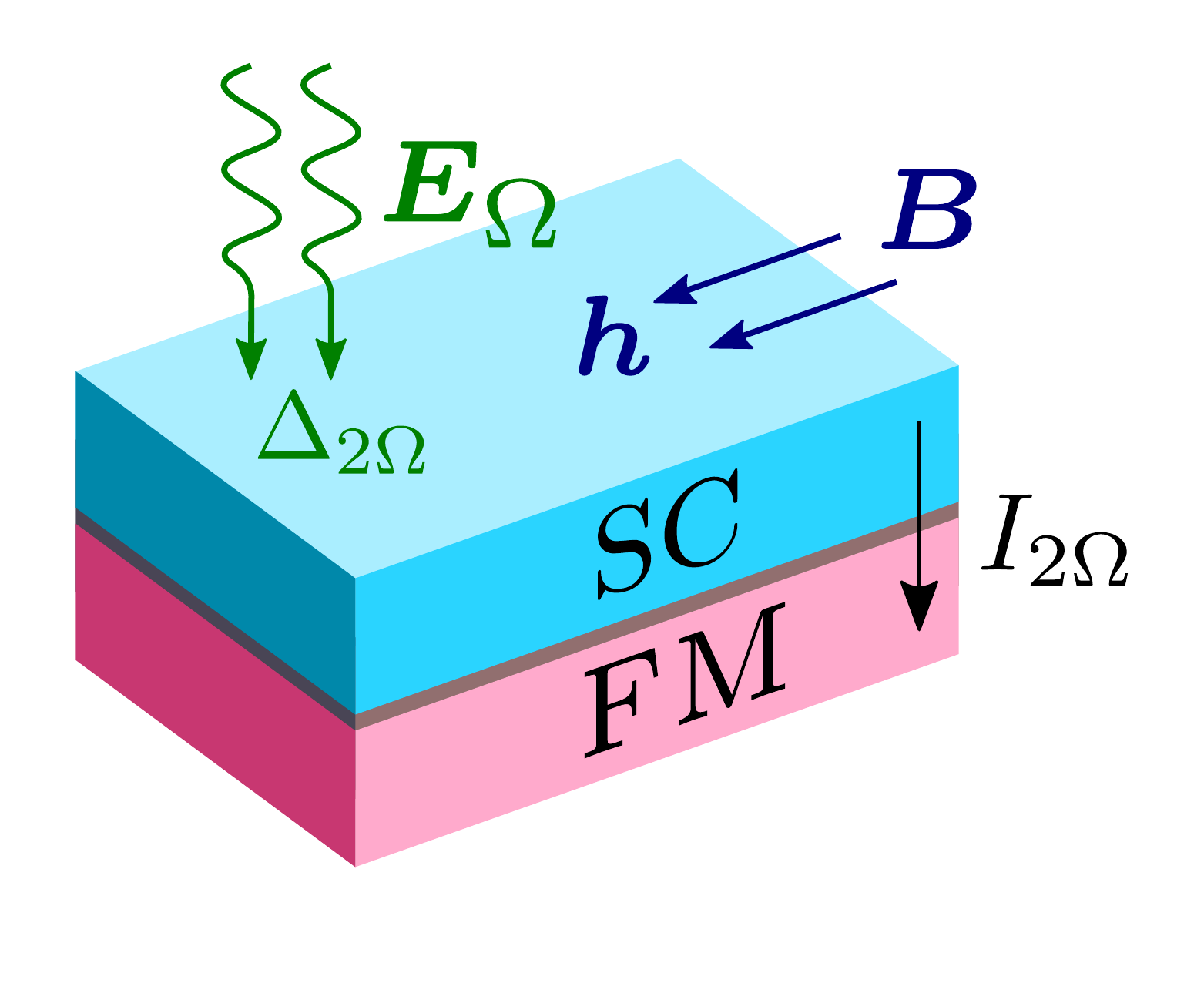}
\includegraphics[width=.45\columnwidth]{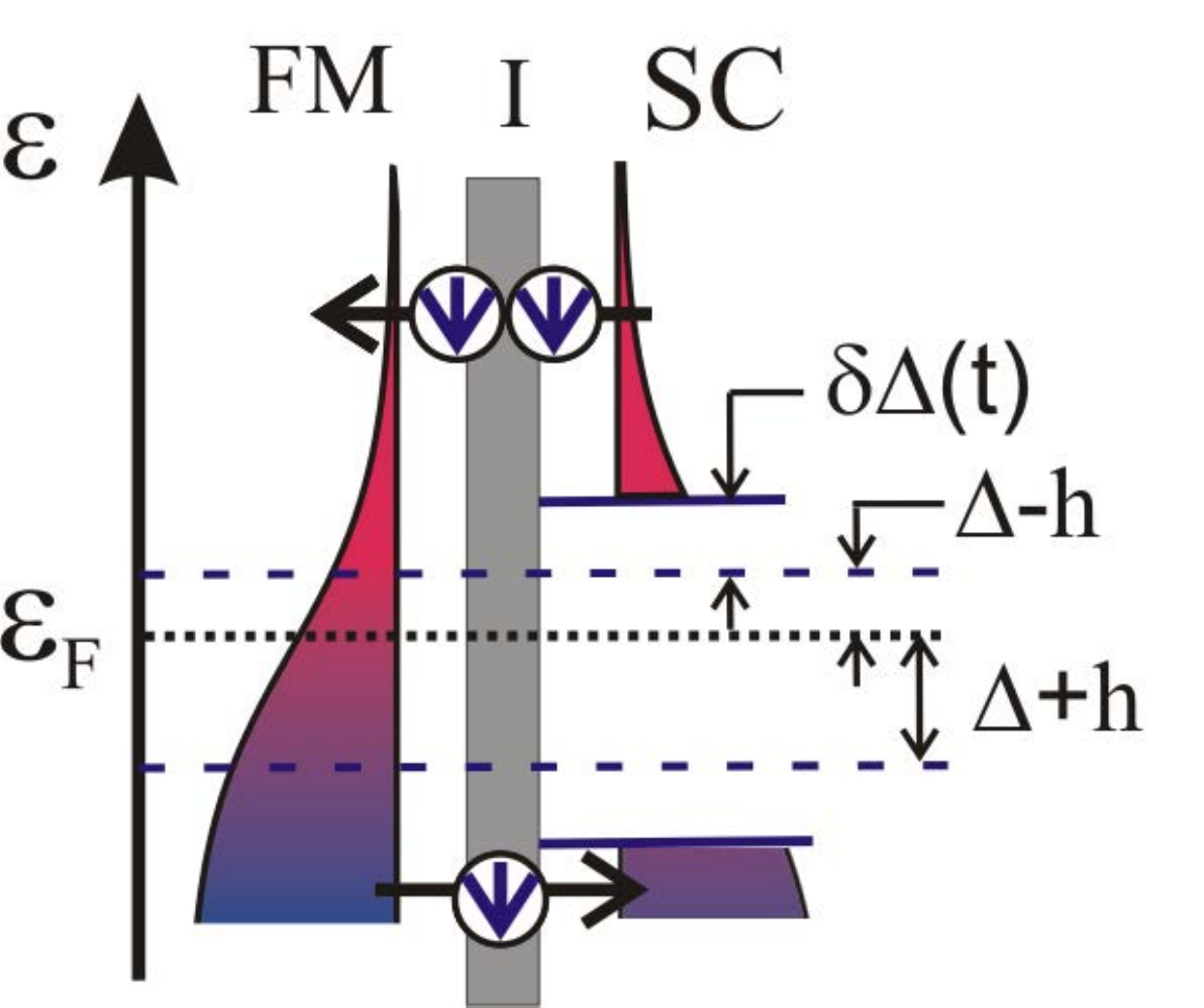}
\caption{\label{fig:schematic} 
(a) Setup of the superconductor (SC) / ferromagnet (FM) structure. The exchange field $\bm h$ is induced by an external magnetic field $\bm B$.
   The double-frequency gap
  modulation can be excited  by the external electromagnetic
  irradiation $\Delta_{2\Omega} \propto E_\Omega^2$ and is enhanced due to the
  coupling to the Higgs mode.
   (b) Sketch of the current generation by the
  time-dependent order parameter amplitude $\Delta (t)$. The shift of the gap amplitude $\delta \Delta (t)$
  pushes occupied quasiparticle states up or down in energy.  Due to the Zeeman shift $h$
  the perturbation of the distribution function is asymmetric with
  respect to the Fermi level $\varepsilon_F$. This results in net spin and charge currents flowing into the attached ferromagnetic electrode (FM) through the tunnel barrier (I).
   }
\end{figure}

{\it Setup}.  The considered setup is shown in Fig.~\ref{fig:schematic}(a). 
 It consists of a superconducting (SC) film placed in contact with a
 ferromagnetic (FM) material. 
 An effective spin splitting field $\bm h$ in SC is induced by an external in-plane magnetic field. Alternatively, $\bm h$ could be induced by the proximity to a ferromagnetic insulator \cite{PhysRevLett.86.3140,
PhysRevB.38.8823, PhysRevB.38.4504,PhysRevB.80.184511,
Eschrig_2015}.
 The system is exposed to external irradiation $E_\Omega e^{i\Omega t} $
 which generates a time-dependent perturbation of the order parameter amplitude $\delta \Delta (t) = \Delta_{2\Omega} e^{2i\Omega t}$ through the second-order non-linearity  $\Delta_{2\Omega} \propto E_\Omega^2$ \cite{Gorkov1968,Gorkov1969}.

 As shown in Fig.~\ref{fig:schematic}(b), the
 time-dependent gap function in the superconducting film creates a non-equilibrium state. 
 Due to the Zeeman shift $h$ this state is non-symmetric with respect to the Fermi level $\varepsilon_F$ and therefore produces spin current through the tunnel barrier between the SC and the adjacent normal metal. This qualitative picture is based on the time-dependent energy spectrum 
 $E= \sqrt{\xi_p^2 + \Delta(t)^2} + \sigma h$ with $\sigma =\pm 1$ for  
 spin-up/down Bogolubov quasiparticles respectively, where $\xi_p$ is the kinetic energy counted from the Fermi level $\varepsilon_F$.  
 
 For a slowly time-dependent order parameter  we find the spin current
 \begin{align} \label{Eq:SpinCurrentSimple}
     I_s (t)= \frac{\kappa}{\Gamma} \dot{\Delta} \frac{d}{d\Delta} (N_+ - N_-),
 \end{align}
 where $\kappa$ is the effective barrier transparency,   $N_\sigma (\Delta) $ is the equilibrium number of thermally excited quasiparticles in the spin-up/down subbands for the given order parameter $\Delta$.  
 The spin current is determined by the energy relaxation Dynes parameter \cite{dynes84} $\Gamma$ and the result \eqref{Eq:SpinCurrentSimple} is shown  to be valid \cite{SupplementaryMaterial} for low frequencies  $\Omega \ll \Gamma$. 
  Expression  \eqref{Eq:SpinCurrentSimple} allows for the cartoon interpretation in terms of the semiconductor model in Fig.~\ref{fig:schematic}(b).  
 However, for the most interesting case when the frequency of $\Delta
 (t)\propto e^{2i\Omega t}$ oscillation is comparable to the gap
 $\Omega \sim \Delta$ and 
 hence is coupled to the Higgs mode
 \cite{Volkov1973,  
 Kulik1981,Barankov2004,Barankov2006}, the picture becomes more complicated. The exact expression for the
 time-dependent spin current valid for all frequencies is derived in
 this Letter [see Eq.~\eqref{Eq:SpinCurrent}].

{\it Second harmonic generation.} Spin currents generated by
the Higgs mode can be detected
 in various ways. The most common approaches for spin current detection are based on the inverse spin Hall effect and spin-filtering systems. Here we rely on the latter possibility which can be achieved by taking into account the spin-dependent transmission probability of the electrons through the SC/FM interface.
 In the setup
 shown in Fig.~\ref{fig:schematic}(a) the spin current is effectively converted to the charge current while passing through the spin-filtering barrier characterized by the polarization vector $\bm P$. The time-dependent charge current induced in this way by the order parameter amplitude oscillation  is therefore qualitatively given by $I (t)\propto \bm P\cdot \bm I_s(t)$,
 which results in the estimate $I (t)\propto (\bm P\cdot \bm h) \partial_t \Delta$. Modulation of the order parameter amplitude
 can be  induced for example by an external irradiation
 \cite{Gorkov1968,Gorkov1969}, $\Delta(t) \propto \bm A^2(t)$, where
 $\bm A(t)$ is the vector potential of the external field.  Hence this
 charge current $I (t)\propto (\bm P\cdot \bm h) \partial_t \bm A^2$, being quadratic in the vector potential, demonstrates the second harmonic generation (SHG) controlled by the superconducting order parameter.

  Despite large attention to the non-linear effects in superconductors,  SHG  has not been obtained before \footnote{Here we exclude the trivial SHG generation which results from the third-order nonlinearity when both the oscillating and constant fields are applied}.
     Hence only the third-harmonic generation has been studied in superconductors \cite{Gorkov1968,Gorkov1969,PhysRevLett.37.930,Matsunaga1145,Giorgianni2019}. 
 We show below that such kind of SHG is not prohibited by the generic
 symmetries of the problem. However it is eliminated by the approximate symmetry of Fermi surface systems, made exact in the 
 widely used quasiclassical approximation \cite{Silaev2017,quasiclassicalnote}.
 For the non-stationary charge current 
generated by the time-dependent vector potential this symmetry yields
$I (\bm A,\Delta) = - I (-\bm A,\Delta^*) $. Further, in the
  absence of supercurrent or external orbital fields we can assume the order parameter to be real $\Delta=\Delta^*$. 
Then even the broken inversion symmetry near  surfaces does not help
to produce SHG in superconducting systems in contrast to the normal
metal counterpart of this effect, and the Higgs mode cannot be measured with this technique.

The particle-hole symmetry is broken to a large extent in SC/FM systems leading to large thermoelectric  \cite{PhysRevLett.110.047002,ozaeta2014predicted,RevModPhys.90.041001} and anomalous Josephson effects \cite{Silaev2017}. 
As shown explicitly below for real $\Delta$
the tunnel charge current through the
spin-polarized barrier satisfies the symmetry   
\begin{align} 
 \label{Eq:jtunnelsymm}
  I (\bm A,  \bm h, \bm P ) = 
 - I (-\bm A,  \bm h, - \bm P).
 \end{align}
 In this case SHG is possible as can be seen from the expression for
 the  tunnel current (\ref{Eq:jtunnelsymm}): Due to the requirement of the sign flip of
 $\bm P$ in the time reversal transformation, 
 there is no longer a symmetry with respect to the mere flipping of
 the vector potential $\bm I (\bm A) \neq - \bm I (-\bm A)$.
 Hence for the ac external field $\bm A_\Omega e^{i\Omega t}$, Eq.~(\ref{Eq:jtunnelsymm}) allows for the double-frequency charge current component $I_{2\Omega} e^{i\Omega t}$ 
 with the amplitude $I_{2\Omega}\propto  |\Delta|^2  A^2_{\Omega}  (\bm P\cdot\bm h) $ 
 as well as the dc tunnel current \cite{PhysRevB.93.014512} $I_{dc} \propto |\Delta|^2 A_{\Omega} A_{-\Omega}  (\bm P\cdot \bm h) $.

 Below we explicitly demonstrate the existence of the double-frequency spin and charge currents in SC/FM tunnel junctions subject to the external electromagnetic irradiation. 
 We show that in general there are two contributions to such spin and
 charge SHG effect. One comes from the direct coupling of electrons in the superconductor to the vector potential. The other is induced by the order parameter amplitude modulation which in turn is excited by the electromagnetic irradiation. 
   
 {\it Spin-polarized tunneling.}
 We model the SC/FM junction using the tunneling  Hamiltonian approach \cite{PhysRevLett.10.486,
PhysRevLett.9.147} which has been extensively used to study both ac
and dc tunnel currents \cite{PhysRevB.30.6419, PhysRevB.11.3329,
  PhysRev.147.255,PhysRevLett.10.486},
 \begin{align}
 & H_T  = \sum_{kk^\prime\alpha} 
  \check A_{k\alpha}^\dagger (\check\Gamma \hat B_{k^\prime})_\alpha + h.c. \\
 & \hat\Gamma = {\cal T}\hat\tau_3 + {\cal U} 
 (\bm m\cdot \hat{\bm\sigma}) .
 \end{align}
 Here $\hat A_{k\alpha}$ ($\hat B_{k\alpha}$) annihilates an electron with momentum $k$ and spin $\alpha$ in the superconductor (ferromagnet), the unit vector $\bm m$ defines the spin quantization axis of the barrier, $\hat\tau_k$ and $\hat\sigma_k$ are the Pauli matrices in Nambu and spin spaces, respectively, and $\cal U$ and $\cal T$ are the spin-independent and spin-dependent matrix elements of the tunneling Hamiltonian \cite{bergeret2012electronic}. 
  The matrix tunneling current through the spin-polarized barriers
  can be expressed through momentum-averaged   Green functions  in the
  superconducting and ferromagnetic electrodes, 
  $\nu_S \hat g_S = \hat\tau_3 \sum_{k} 
  \langle {\,\timeorder}  \hat A_k (t) \hat A^\dagger_{k} (t^\prime) \rangle$ and $\nu_F \hat g_F = \hat\tau_3 \sum_{k} 
  \langle {\,\timeorder}  \hat B_k (t) \hat B^\dagger_{k} (t^\prime)
  \rangle$, respectively. 
   Here $t, \; t^\prime$
  are imaginary times, $\timeorder$ is the time-ordering operator, $\nu_{S/F}$ are the normal metal densities
  of states on the two sides of the junction. For simplicity we
  assume momentum-independent tunneling coefficients \cite{bergeret2012electronic,
Bergeret2012a}. 
  The time-dependent tunneling current 
  for the general non-equilibrium state in the electrodes \cite{SupplementaryMaterial} reads 
 \begin{align} \label{Eq:TunnelCurrent}
   \hat I (t) = i \frac{\nu_S \nu_F}{2}
  [  {\hat g}_S\circ   
  (\hat\Gamma
  {\hat g}_F\hat\Gamma)
  - (\hat\Gamma{\hat g}_F 
  \hat\Gamma)\circ
  {\hat g}_S
  ], 
   \end{align}
   where ${\hat g}_{S(F)} $ is the quasiclassical GF in the SC  (FM) electrode and $\circ$ denotes time convolution. 
   The  overall tunnel current amplitude is determined   by $\kappa=
  \nu_S \nu_F ({\cal T}^2 + {\cal U}^2 )$ and the effective spin-filtering polarization is  $\bm P =  2 {\cal T} {\cal U} \bm m/({\cal T}^2 + {\cal U}^2 )$. 
  Tracing the general expression with appropriate Pauli matrices we  extract the charge current $I =e {\rm Tr}( \hat\tau_3 \hat I)$ and the spin current $\bm I_s = {\rm Tr}( \hat{\bm \sigma} \hat I)$, respectively. 
  
  We assume that the electrodes are described by 
  the time-dependent quasiclassical Usadel theory 
 and include only lowest order [$o(\kappa)$] corrections
    from tunneling \cite{1906.02751}. 
 This conventional approximation allows for the spin currents driven by the Higgs mode and external field even with a non-ferromagnetic barrier, that is at $P=0$.
 However, the  direct coupling between the Higgs mode and the charge current is prohibited by the particle-hole symmetry. As shown in the Supplementary Material \cite{SupplementaryMaterial} for the solutions of Usadel equation this symmetry yields 
  \begin{align} \label{Eq:gSymm}
     \check g(\bm A, \bm h, \Delta) = - \hat\tau_1 \check g(- \bm A, \bm h, \Delta^*) \hat\tau_1 .
\end{align}
Here the off-diagonal Nambu space Pauli matrix $\hat\tau_1$ interchanges the particle and hole blocks in the Hamiltonian \cite{Silaev2017}.  
 This symmetry is broken by the spin polarization of tunneling $P\neq 0$ so that
 the transformation (\ref{Eq:gSymm}) applied to the general tunnel current yields the relation (\ref{Eq:jtunnelsymm}) which allows for the finite charge current. 

  Further we assume that the SC electrode  is driven out of equilibrium  by the external irradiation. It generates the second-harmonic perturbation of the GF and tunnel current  
  \begin{align}
   &    \hat g_{S} (t,t^\prime)=   T  \sum_\omega \hat g_{S} (\omega_+,\omega_-)
  e^{i\omega_+ t - i \omega_- t^\prime}
  \\
  \label{Eq:QuasiclassicalImaginary1}
  & \hat I ({2\Omega}) = iT \frac{\nu_S \nu_F}{2}  \sum_\omega   
  \hat\Gamma \hat g_{S} \hat\Gamma 
 {\hat g}_0(\omega_-)
  - 
  {\hat g}_0(\omega_+)
 \hat\Gamma \hat g_{S}  \hat\Gamma,
  \end{align}
  where $\omega_\pm = \omega\pm \Omega $ are the fermionic Matsubara
 frequencies shifted by the frequency $\Omega$ of the external field. 
   We denote  $\hat g_S = \hat g_{S}(\omega_+,\omega_-) $  and assume
  that the ferromagnet is in the equilibrium state determined  by the
  GF $g_F(\omega) = g_0(\omega) \equiv {\rm sign}(\omega) \hat\tau_3$.
  There are two qualitatively different terms in  the
    non-equilibrium GF $\hat g_S = \hat g_\Delta + \hat g_{AA}$. The
    first one is generated by the time-dependent order parameter
    whereas the second term is generated by the direct coupling to the external field. Below we discuss the corresponding contributions to the tunnel current and coupling to 
 the Higgs mode.
 
  \begin{figure}
\includegraphics[width=1.0\columnwidth]{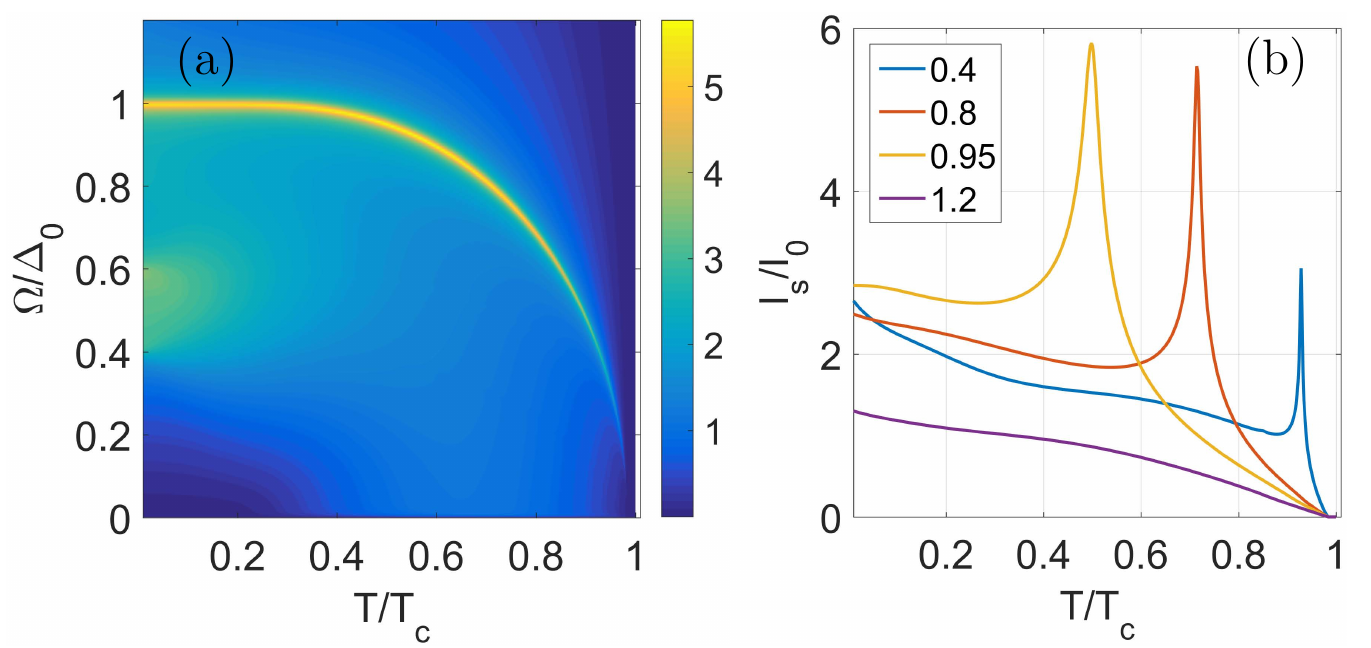}
\caption{\label{fig:IOmTemp}  Amplitude of double-frequency spin current  $I_s(2\Omega) e^{2i\Omega t}$ driven through the SC/FM tunnel junction by an external field $A_\Omega e^{i\Omega t} $. 
 The current is  normalized to 
 $I_0 = \kappa T_c D(eA_\Omega/c)^2$, where 
 $T_c$ is the critical temperature. The sharp maximum at $\Omega=\Delta(T)$ corresponds to the resonant excitation of the Higgs mode. (a) $I_s(\Omega, T)$,
(b) $I_s(T)$ at different frequencies $\Omega/\Delta_0=0.4;\;0.8;\;0.95;\;1.2$. 
The exchange field is $h=0.2\Delta_0$ and Dynes parameter $\Gamma =0.005 \Delta_0$. 
The peaks are at temperatures determined by $\Omega=\Delta(T)$.
 }
 \end{figure}
 
  {\it Higgs mode contribution.}
 First, let us discuss the term $\hat g_\Delta$ which is generated by the  time-dependent order parameter amplitude  $\Delta_{2\Omega}
 e^{2i\Omega t}$. 
  The correction to the GF driven by the time-dependent order parameter field is given by \cite{SupplementaryMaterial}
        \begin{align} \label{Eq:g13Delta}
   \hat g_{\Delta}  =   \Delta_{2\Omega}
   \frac{ \hat g_0(\omega_+)\hat\tau_2 \hat g_0(\omega_-)-\hat\tau_2}
   {s (\omega_+) + s(\omega_-) },
   \end{align}  
   where the Nambu-space Pauli matrix  $\hat\tau_2$
   is the vertex describing the coupling of electrons to the order
   parameter field. 
   In this expression the denominator contains 
   $s(\omega)=[(\omega+i \sigma h)^2 + \Delta^2]^{1/2}$, where $\sigma=+/- $ corresponds to spin-up/down subbands. 
     Substituting Eq.~\eqref{Eq:g13Delta} to the general matrix current
  (\ref{Eq:TunnelCurrent}) and using analytical continuation
  \cite{KopninBook,SupplementaryMaterial} we obtain  the amplitude of real-frequency spin current $I_s(2\Omega) e^{2i\Omega t}$  
     \begin{align} \label{Eq:SpinCurrent}
  & I_s = i  \kappa \Delta_{2\Omega}  \Delta \sum_\sigma \sigma \int \frac{d\varepsilon}{4\pi} 
    \frac{\varepsilon [n(\varepsilon_+) - n (\varepsilon_-)] }{s^R_+ s^A_-(s^R_+ + s^A_-)},
  \end{align}  
  where $n(\varepsilon)$ is the equilibrium distribution function, $\varepsilon_\pm = \varepsilon \pm \Omega + \sigma h$ and 
    $ s^{R,A}  = - i \sqrt{ (\varepsilon \pm i\Gamma)^2- \Delta^2} $. 
  
   In the low-frequency limit $\Omega \ll \Gamma$ we restore Eq.~\eqref{Eq:SpinCurrentSimple} when the spin current is driven by the adiabatic time dependence of $\Delta$ in accordance with the qualitative picture shown schematically in Fig.~\ref{fig:schematic}(b).
   The numbers of thermally excited states are 
   $ N_\sigma = \int d\xi_p n(E_\sigma(\xi_p,\Delta))
  $ where $E_\sigma = \sqrt{\xi_p^2 + \Delta^2} + \sigma h$ is the spin-splitted  spectrum of Bogolubov quasiparticles.
   
   In the presence of the Higgs mode, that is the slowly
     decaying oscillations of the order parameter 
     $\Delta (t)$
       \cite{Volkov1973,Barankov2004},  the spin current is given by the sum of the corresponding Fourier components with the amplitudes given by  (\ref{Eq:SpinCurrent}). 
       As a result of Eq.~\eqref{Eq:SpinCurrent} we get slowly-decaying oscillations of the spin
    current $I_s (t)$ which can be
    measured using electrical probes after the superconductor is
    initially driven into a non-equilibrium state by a field pulse.

  {\it Coupling to an external field.}
  The Higgs mode can also be revealed by the spin current if the superconductor is driven out of equilibrium by a continuous wave irradiation
 as shown schematically in Fig.~\ref{fig:schematic}(a).
   The second-order direct coupling to the external field is determined by the GF perturbation 
   $\hat g_{AA}\propto A_\Omega^2$. 
        In the dirty limit this term 
 can be found from the Usadel equation as described in the Supplementary Material \cite{SupplementaryMaterial}
     \begin{align}
  & \hat g_{AA} =  D\left(\frac{eA_\Omega}{c}\right)^2 \frac{\hat\tau_3\hat g_0 (\omega)\hat\tau_3 - \hat g_0 (\omega_+)\hat\tau_3 
  \hat g_0 (\omega) \hat\tau_3 \hat g_0(\omega_-)}{s(\omega_+) + s(\omega_-)},\label{eq:gAA1}
   \end{align}   
 where $D$ is the diffusion coefficient. 
 
 This coupling to the external field has a twofold effect. First it directly generates second-harmonic  spin and charge currents. Besides that it generates the time-dependent component of the order parameter 
 according to the self-consistency equation. 
 The bare amplitude of the order parameter perturbation  $F_\Delta e^{2i\Omega t}$ is given by 
  $F_\Delta= - \lambda T \sum_\omega
    {\rm Tr} [\hat\tau_2 \hat g_{AA} ] $, where $\lambda$
    is the coupling constant of superconductivity
    and the Pauli matrix $\hat \tau_2$ corresponds to the  superconducting amplitude vertex.
          
          The resonant excitation of the total order parameter
          amplitude is determined by the equation with polarization
          corrections 
          $\Delta_{2\Omega} = F_\Delta  + \Pi(2\Omega) \Delta_{2\Omega}$, 
           where $\Pi(\Omega)$ is the order parameter polarization operator \cite{PhysRevB.92.064508, Cea2016}.
          It has  the simple solution
          $\Delta_{2\Omega} = F_\Delta /[1-\Pi (2\Omega)]$, that 
    yields the resonance condition for 
      $\Omega=\Delta$ when the denominator satisfies $1-\Pi(2\Delta) = 0 +
      o(\sqrt{\Gamma}) $. Hence the maximal amplitude of the spin
      current is determined by the broadening parameter $\Gamma$,
       leading to a sharp peak in $I_s(\Omega,T)$ for $\Omega \approx \Delta(T)$.
    This behaviour of the spin current is illustrated in Fig.~\ref{fig:IOmTemp}. 
    The charge current appearing in the case of a
    finite spin polarization of the barrier is 
    given by  $I= e \bm P \cdot \bm I_s$, where the 
    spin current vector is $\bm I_s = I_s \bm h /h $. 
          
  \begin{figure}
\includegraphics[width=1.0\columnwidth]{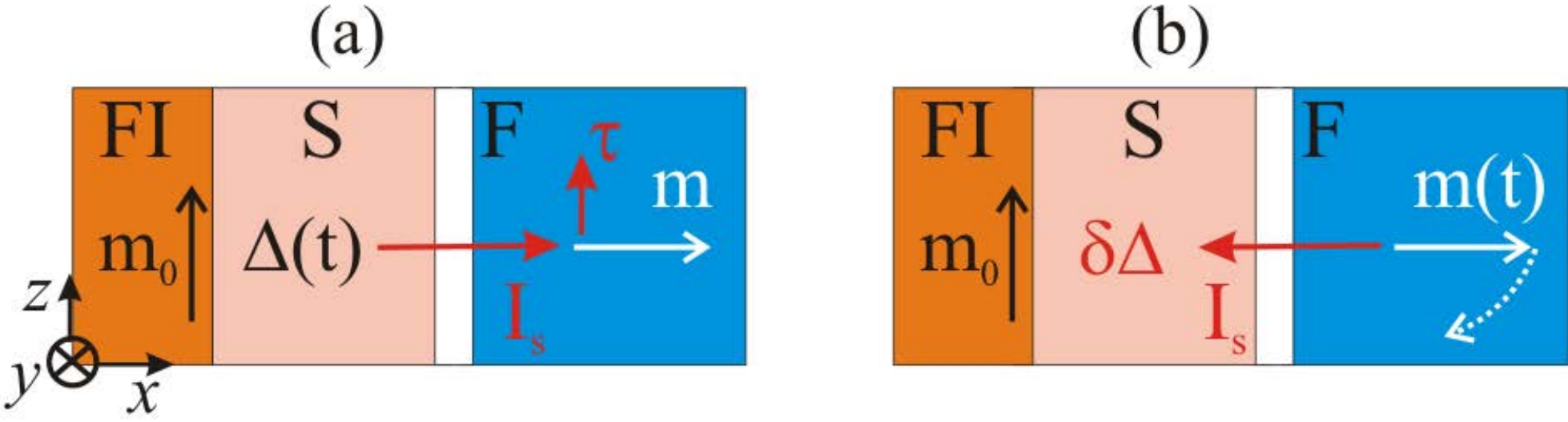}
\caption{\label{fig:SpinTorque}
The setups to study transverse spin currents coupled to the Higgs mode in the superconductor (S). The exchange field in S is generated by the adjacent ferromagnetic insulator (FI) with magnetization $\bm m_0$. 
(a) The spin torque $\bm \tau$ is generated in an adjacent ferromagnet (F) with non-collinear magnetization $\bm m \nparallel \bm m_0$. 
(b) Magnetization precession $\bm m(t)$ induces the spin current $\bm
I_s$ and the spin battery effect leading to a perturbation of the order parameter amplitude $\delta \Delta$.
}
\end{figure}

{\em Spin torques.}
 If the exchange field  $\bm h$ in the SC is non-collinear with the magnetization $\bm m $ in the FM,
 the Higgs mode generates a spin torque acting on $\bm m$.  The generic system which can realize this configuration is
shown in Fig.~\ref{fig:SpinTorque}(a). Here the exchange field $\bm
h\parallel \bm m_0$ is created by the ferromagnetic insulator (FI)
layer with a fixed magnetic moment $\bm m_0$ \cite{RevModPhys.90.041001}. 
   
   First let us discuss the STT generated by the Higgs mode as shown schematically in Fig.~\ref{fig:SpinTorque}(a).
   The polarization of the non-equilibrium  spin current $\bm I_s$ in
   the superconductor electrode depends only on the direction of the
   exchange field $\bm h$ and is not sensitive to the magnetic moment
   $\bm m$ in the adjacent FM. This is in contrast to the
   equilibrium components of the spin current which exist in such
systems with non-collinear magnetic moments
   \cite{Silaev2015a} even without the external drive, and are
   proportional to $\propto \bm h\times \bm m$. 
    Assuming that the transverse component of the spin current is
      absorbed in the ferromagnet \cite{RevModPhys.77.1375, SLONCZEWSKI1996L1, PhysRevLett.84.2481, PhysRevB.62.12317, PhysRevB.66.014407}
      we obtain the STT  $\bm
      \tau = I_s \bm h_\perp/h$ where $\bm h_\perp = \bm h - \bm m
      (\bm m \cdot \bm h)$ is the perpendicular component of the exchange field.

   This effect can be viewed as the Higgs-mode mediated transfer of
   the spin angular momentum from the FI to the metallic ferromagnet shown in Fig.~\ref{fig:SpinTorque}(a).
        Oscillating STT generated by the order parameter amplitude mode can excite the ferromagnetic resonance (FMR) in the attached ferromagnet. Hybridization of FMR and Higgs resonances should show up as the avoided crossing of the peaks in the 
      second-harmonic response of the systems shown in Fig.\ref{fig:SpinTorque}a. Such experiment will directly  demonstrate the  dynamical coupling of the magnetic and superconducting orders. Modification of FMR resonance linewidth by superconducting correlations in FM/SC structures has been observed recently\cite{PhysRevLett.100.047002,Jeon2018}.

   Experimentally  this effect can be realized using nanomagnets \cite{Yakushiji2005,
Krause2007, Loth2010, 
PhysRevB.93.064407} because in these small-sized systems 
  one can achieve larger coupling between FMR and Higgs mode. The other alternative is to use  molecular magnets coupled to the superconductor and observe the spin currents using macroscopic quantum tunnelling effect\cite{Thomas1996}. 
    
   The reciprocal effect which results in the generation of the gap function amplitude 
   perturbation $\delta\Delta$ 
   by the magnetic precession is 
 shown in
   Fig.~\ref{fig:SpinTorque}(b).
   To demonstrate the possibility to induce  $\delta\Delta (t)$ by magnetic precession we  assume that the  spin current
   $\bm I_s \propto \bm m\times \dot{\bm m}  $  
   is pumped by the time-dependent magnetization $\bm m(t)$ in the FM \cite{RevModPhys.77.1375}. 
   This spin current has a longitudinal component 
   $\bm I_s \parallel \bm h$ which 
   generates a spin accumulation in the superconductor. 
   For a low frequency of the magnetic
     precession this  effect can be described by the  spin-dependent chemical potential shift $\mu_s$ in the SC. In combination with the spin-splitting field $\bm h$ the spin accumulation results in a perturbation of the gap function amplitude
     \cite{PhysRevB.93.014512, PhysRevB.96.104515}
    \begin{align}\label{Eq:DeltaIs}
        \delta \Delta =\frac{\lambda\Delta}{1-\Pi}  \mu_s \partial_\Delta (N_+ -N_-),  
    \end{align}
    where $1-  \Pi\propto  \lambda $ is the low-frequency asymptotic of the polarization operator. This expression demonstrates the possibility to couple the order parameter amplitude with the magnetization dynamics. Thus the higher-frequency 
    magnetization precession with $\Omega \sim \Delta$ generates  the
    Higgs mode in the superconductor with a spin-splitting field.

{\em Conclusions.} 
In this Letter, we have demonstrated that spin currents can be effectively generated by the collective amplitude modes of the superconducting order parameter.
Owing to the fact that  the Higgs mode  can be generated by the external irradiation \cite{PhysRevB.99.224511,
PhysRevB.99.224510},  our result paves the way for conceptually new direction of superconducting optospintronics -- the study of spin currents and spin torques generated by light interacting with superconducting materials. 

We have suggested a detection scheme for the Higgs mode based on
measuring resonant electric signals, either the charge current or
voltage generated across the spin-polarized tunnel junction by the
external field. Because these signals appear at the doubled frequency
of the  external field, our setup introduces the system featuring the SHG effect controlled by superconductivity. The suggested SHG effect can be studied  using optical or microwave detectors. Being sensitive to the 
magnitude of the spin splitting field and the quality of the spin-polarized
barriers, this effect provides a tool for the diagnostics of large-area
SC/FM junctions suggested 
to be  used as a new  platform for
the fabrication of radiation sensors \cite{heikkila2018thermoelectric}.  
    The ac tunneling current $I_{2\Omega}$ can be detected using electrical probes allowing for electrical detection of the Higgs mode in superconductors. 
 
 {\em Acknowledgements} 
 This work was supported by the Academy of Finland (Projects
 No. 297439 and 317118), Jenny and Antti Wihuri Foundation,  Russian Science Foundation (Grant No. 19-19-00594) and from  the  European  Union’s  Horizon  2020  research  and  innovation programme  under  grant  agreement  No  800923  (SUPERTED).

\section{Supplementary}
 
   \subsection{Tunnel current}
  We model SC/I/FM junction by the tunneling Hamiltonian: 
 \begin{align}
 & H_T  = \sum_{kk^\prime\alpha} 
  \check A_{k\alpha}^\dagger (\check\Gamma \hat B_{k^\prime})_\alpha + h.c. \\
 & \hat\Gamma = {\cal T}\tau_3 + {\cal U} 
 (\bm m \bm\sigma) 
 \end{align}
 where the unit vector $\bm m$ is the spin-filtering axis of the  barrier. 
 This model describes spin-dependent tunnelling
 through the SC/FM interface \cite{bergeret2012electronic}.  
 We calculate the tunneling current as a function of the time on the  contour running along the imaginary time axis from $0$ to 
 $\beta = 1/T$. 
The matrix tunneling current in terms of the 
 imaginary-time functions reads
 \begin{align}\label{Eq:CurrentKeldysh}
 \hat I (t) = \frac{i}{2} \sum_k \left[ 
 \partial_t \hat G_S (t,t^\prime,k, k) +  
 \partial_{t^\prime} 
 \hat G_S (t,t^\prime,k,k)\right]|_{t=t^\prime}. \end{align}
  To find the perturbation we consider the 
  contour-ordered GF 
 \begin{align}
 \hat G_S (t,t^\prime,k,k^\prime) = \langle\,\contourorder \hat S 
 \hat A_k (t) \hat A^\dagger_{k^\prime} (t^\prime) \rangle,
 \end{align}
 where $\contourorder$ is the contour-ordering operator and
 \begin{align}
 \hat S \approx 1 - \int_0^\beta dt_c H_T (t_c).
 \end{align}
 
  In the interaction representation with respect to the tunneling  
 Hamiltonian the equation of motion is
 \begin{align}\label{Eq:Motion}
 \partial_t \check A_{k} = [\check A_{k},H_T] = 
  \sum_{k^\prime} 
  (\check\Gamma_{k k^\prime} \hat B_{k^\prime})
  \end{align}

  Using the equation of motion we get 
  \begin{align}
  & \partial_t \hat G_S (t,t^\prime,\bm k,\bm k) = 
  \sum_q \langle\,\contourorder \hat S 
  [\hat\Gamma_{kq} \check{B}_{q}(t)]
  \check{A}^\dagger_{k\beta}(t^\prime) \rangle 
  \approx 
  \\
  & - \sum_{q} 
  \langle\,\contourorder \int_0^\beta dt_c H_T (t_c)
  \hat\Gamma_{kq} \check{B}_{q}(t) 
  \check{A}^\dagger_{k}(t^\prime) \rangle  = 
  \\ \nonumber
  & - \sum_{k_1,k^\prime_1,q}
  \langle\,\contourorder \int_0^\beta dt_c
  [ \check{B}^\dagger_{k^\prime_1}(t_c)
  \hat\Gamma^\dagger_{k_1 k^\prime_1}]
  \check{A}_{k_1}(t_c)
  [ \hat\Gamma_{kq}\check{B}_{q}(t)]
  \check{A}^\dagger_{k}(t^\prime) \rangle 
  = 
  \\ \nonumber
  & -\sum_{k_1,k^\prime_1,q}\int_0^\beta dt_c
  \{ \hat\Gamma
  \langle\,\contourorder 
  \check{B}_{q}(t) 
  \check{B}^\dagger_{k^\prime_1}(t_c) \rangle
  \hat\Gamma(t_c) \}
  \langle\,\contourorder
  \check{A}_{k_1}(t_c)
  \check{A}^\dagger_{k}(t^\prime)
  \rangle =
  \\ \nonumber
  & -\sum_{k_1,k_1^\prime,q}\int_0^\beta dt_c 
   \hat\Gamma_{kq} {\hat G}_F(t,t_c,q,\bm k_1) 
   \hat\Gamma_{\bm k^\prime_1 k_1}  
  {\hat G}_S (t_c,t^\prime,\bm k^\prime_1,\bm k)
  \end{align}  
  and  
  \begin{align}
  & \partial_{t^\prime} \hat G_S (t,t^\prime,\bm k,\bm k) = 
  \sum_q \langle\,\contourorder \hat S 
  \check{A}_{k}(t) [\hat\Gamma_{kq} 
  \check{B}_{q}(t^\prime)]^ \dagger \rangle 
  \approx 
  \\ \nonumber
   & - \sum_{q} 
  \langle\,\contourorder \int_0^\beta dt_c H_T (t_c)
  \check{A}_{k}(t)
  [\hat\Gamma_{kq} \check{B}_{q}(t^\prime)]^\dagger
  \rangle  = 
  \\ \nonumber
  & - \sum_{k_1,k^\prime_1,q}
  \langle\,\contourorder \int_0^\beta dt_c 
  [ \hat\Gamma_{k_1 k_1^\prime}\check{B}_{k^\prime_1}(t_c)]
  \check{A}^\dagger_{k_1}(t_c)
  \check{A}_{k}(t)
  [ \check{B}^\dagger_{q}(t^\prime)\hat\Gamma^\dagger_{qk}]
  \rangle = 
  \\ \nonumber
  & \sum_{k_1,k^\prime_1,q}\int_0^\beta dt_c 
  \langle\,\contourorder
  \check{A}_{k}(t)
  \check{A}^\dagger_{k_1}(t_c)
  \rangle  
  \{ \hat\Gamma_{k_1 k^\prime_1}
  \langle\,\contourorder
  \check{B}_{k^\prime_1}(t_c) 
  \check{B}^\dagger_{q}(t^\prime) \rangle
  \hat\Gamma_{qk} \}
   = 
   \\ \nonumber
  & \sum_{k_1,k^\prime_1,q} 
  \int_0^\beta dt_c 
  {\hat G}_S (t,t_c,\bm k, \bm k_1)
  \hat\Gamma_{k_1 \bm k_1^\prime} 
  {\hat G}_F(t_c,t^\prime,\bm k_1^\prime, q) \hat\Gamma_{q,k}. 
  \end{align}

 Hence the matrix current is given by 
  \begin{align}
  & \hat I (t) = 
  \frac{i}{2}\sum_{k, k_1,k_1^\prime, q}
   \\ \nonumber
  & \int_0^\beta dt_c \{
   {\hat G}_S(t,t_c,\bm k, \bm k_1)   
  \hat\Gamma_{k_1 \bm k_1^\prime} 
  {\hat G}_F(t_c,t, \bm k_1^\prime, \bm q)\hat\Gamma_{qk}
  \\ \nonumber
  & - \int_0^\beta dt_c 
  \hat\Gamma_{k,q}{\hat G}_F (t,t_c,\bm q, \bm k_1)
  \hat\Gamma_{k_1\bm k_1^\prime}
  {\hat G}_S(t_c,t,\bm k_1^\prime,\bm k) 
  \} .
  \end{align}
  
  We assume that GFs are spatially homogeneous, 
  so that 
  ${\hat G}_F (t,t_c,\bm q, \bm k_1) = 
  \delta_{q,k_1}\hat\Gamma{\hat G}_F (t,t_c,\bm q) $ 
  and the matrix element is momentum-independent
  $\hat \Gamma_{kk^\prime} = \hat \Gamma$. Then we can introduce the quasiclassical functions $ \sum_q {\hat G}_{F,S} (t,t_c,\bm q) = \nu_{F,S}\hat\tau_3 \hat g_{F,S}(t,t_c) $
 to write the current as  
 \begin{align} \label{Eq:ItApp}
  \hat I (t) = i \frac{\nu_S \nu_F}{2}
   [  {\hat g}_S\circ   
  (\hat\tau_3\hat\Gamma
  {\hat g}_F\hat\Gamma\hat\tau_3)
  - (\hat\tau_3\hat\Gamma{\hat g}_F 
  \hat\Gamma\hat\tau_3)\circ
  {\hat g}_S
  ],
  \end{align}
  where the time convolution symbol is defined as $(A\circ B)(t,t_1) = \int_0^\beta dt_c  A (t,t_c)B(t_c,t)$. 
  Taking into account that the normal metal GF ${\hat g}_F $ commutes with $\hat\tau_3$, Eq.~\eqref{Eq:ItApp}
 can be reduced to Eq.~\eqref{Eq:TunnelCurrent} in the main text.

\subsection{Analytical continuation}
   \label{Sec:AnalyticalContinuation}
 In order to find the real-frequency response we need to implement 
 the analytic continuation of Eq.~\eqref{Eq:QuasiclassicalImaginary1}. 
 These second-order responses are obtained by the summation of expressions 
 which depend on the multiple 
 shifted fermionic frequencies such as 
 $g(\omega_1,\omega_2,\omega_3)$. 
 The analytic continuation of the sum by Matsubara frequencies 
 is determined according to the general rule \cite{KopninBook}
  \begin{align} \label{Eq:AnalyticalContinuationGen}
  & T\sum_\omega g(\omega_1,\omega_2,\omega_3)
  \to  
  \\ \nonumber
  & \sum_{l=1}^3 \int \frac{d\varepsilon}{4\pi i}  
  n_0(\varepsilon_l)
  \left[ g(...,  -i \varepsilon^R_l, ...) - 
  g(...,  -i \varepsilon^A_l, ...)  \right],
  \end{align}
 where $n_0(\varepsilon) = \tanh (\varepsilon/2T ) $ is the equilibrium  
 distribution function. In the r.h.s. of (\ref{Eq:AnalyticalContinuationGen}) we substitute in each term 
 $\omega_{k<l} = - i\varepsilon^R_k$ and 
 $\omega_{k>l} = - i\varepsilon^A_k$ for $k=1,2,3$,
  denote 
  $\varepsilon_k = \varepsilon + (2-k)\Omega$ 
  and  $\varepsilon^R = \varepsilon + i\Gamma$, $\varepsilon^A = 
  \varepsilon -i \Gamma$. Here the term with $\Gamma>0$ is added to   
  shift  the integration  contour into the corresponding half-plane.
  At the same time,  $\Gamma$ can be used as the Dynes 
  parameter \cite{dynes84}
  to describe the effect of different depairing mechanisms 
  on spectral functions in the superconductor.
  We implement the analytical continuation in such a way that 
  $
  s(-i\varepsilon^{R,A}) 
  = - i \sqrt{ (\varepsilon^{R,A})^2- \Delta^2}
   $    assuming that the branch cuts run from $(\Delta,\infty)$
   and $(-\infty, -\Delta)$. 
 In the presence of the spin-splitting field
 the energy  in Eq.~\eqref{Eq:AnalyticalContinuationGen} should be shifted to $\varepsilon + \sigma h$,
 where $\sigma= \pm 1 $ is the spin subband index.
 
 Equilibrium GF in the imaginary frequency domain is given by $\hat g_0(\omega ) = (\hat\tau_3 \omega - \hat\tau_2 \Delta)/s(\omega)$.
 The real-frequency continuation reads
   $\hat g^{R,A}_0(\varepsilon )  
 =  (\hat\tau_3 \varepsilon_{R,A} -i \hat\tau_2 \Delta)/\sqrt{ (\varepsilon^{R,A})^2- \Delta^2}$.

 {\it Example.}  %
 To demonstrate the analytical continuation in practice we calculate the spin current driven by the Higgs mode.
 For real frequencies the spin  current obtained from  (\ref{Eq:ItApp}) can be written in terms of the Keldysh component
  \begin{align} \nonumber
 & I_s = \frac{\kappa}{8\pi} \sum_\sigma \sigma
 \int d\varepsilon {\rm Tr} [ \hat g_F (\varepsilon_+) \hat g_S (\varepsilon) - \hat g_S (\varepsilon) \hat g_F (\varepsilon_-)  ]^K  = 
  \\ \label{EqSM:SpinCurrent1}
  & \frac{\kappa}{8\pi} \sum_\sigma \sigma
 \int d\varepsilon [n(\varepsilon_+) - n (\varepsilon_-)] {\rm Tr} [\hat\tau_3 g^a_S],
 \end{align}   
 where $\varepsilon_\pm = \varepsilon + \sigma h \pm \omega$. 
  In deriving (\ref{EqSM:SpinCurrent1}) we used the fact that 
 $\hat g^{R(A)}_F = \pm 1$ do not dependent on energy.
  The anomalous part of the nonequilibrium GF in the superconductor is
 \begin{align} \label{EqSM:ga}
   {g}^a_S = \Delta_{2\Omega} \frac{ g^R (\varepsilon_+)\tau_2 g^A (\varepsilon_-) - 
   \tau_2 }{s^R_+ + s^A_-  }
 \end{align}
   where we denote $s^{R,A}_\pm = s^{R,A} (\varepsilon_\pm)$.
  
   Substituting the solution (\ref{EqSM:ga}) and using  ${\rm Tr} [\tau_3 g^R_+\tau_2 g^A_-] = 2i\Delta_0 \varepsilon/s^R_+ s^A_-$,
 we get 
 \begin{align} \label{Eq:Current1}
  & I_s = i\kappa\Delta_0 \Delta_{2\Omega}\sum_\sigma \sigma \int \frac{d\varepsilon}{4\pi} 
    \frac{(\varepsilon+\sigma h) [n(\varepsilon_+) - n (\varepsilon_-)] }{s^R_+ s^A_-(s^R_+ + s^A_-)} = 
  \\ \nonumber
  &  \frac{i\kappa \Delta_0\Delta_{2\Omega}}{(\omega + i\Gamma)} \int \frac{d\varepsilon}{16\pi}  \sum_\sigma \sigma 
  [n(\varepsilon_+) - n (\varepsilon_-)] 
  \left( \frac{1}{s^R_+} - \frac{1}{s^A_-} \right),
  \end{align}   
  where we use $(s^R_+)^2-(s^A_-)^2 = 4(\varepsilon+\sigma h) (\omega+i\Gamma)$.
  In the low-frequency limit we can substitute 
  $  n(\varepsilon_+) - n (\varepsilon_-) = 2\omega \partial_\varepsilon n $ and ${s^R_+} = - {s^A_+} = - i \sqrt{\varepsilon^2-\Delta^2}$. 
  Then spin current can be written in the simple form  
  \begin{align}
   I_s = \frac{\kappa}{\Gamma} \sum_\sigma \sigma \frac{d}{dt} \int d\xi_p n(E_\sigma(\xi_p,\Delta(t)))
   =
   \frac{\kappa}{\Gamma} \dot{\Delta}\frac{d}{d\Delta} (N_+ - N_-)
  \end{align}
 where $E_\sigma(\xi_p,\Delta(t)) = \sqrt{\xi_p^2 + \Delta^2(t)} + \sigma h$ is the spectrum of Bogolubov quasiparticles shifted by the spin-splitting field $h$. 


\subsection{Description in terms of the time-dependent Usadel equation }
We start by analyzing the symmetries of the
current in a superconductor driven by the time-dependent external field vector potential 
$\bm A (\bm r, t)$, order parameter $\Delta (\bm r, t)$  and exchange field $\bm h (\bm r,t)$. 
Superconductor is described by Usadel equation, which is a  
  diffusion-like equation for the  
 quasiclassical Green functions (GF). 
 In the imaginary time representation it has the form 
 \begin{equation}  \label{Eq:UsadelGen}  
 -i \{\hat\tau_3\partial_\tau, \check g \}_\tau = D\hat\partial_{\bf r}  ( \check g
\circ \hat\partial_{\bf r} \check g) 
 - i [\hat\tau_3\hat H , \check g ]_t 
   \end{equation}
where $D$ is the diffusion constant, 
$\hat H=  \Delta\hat\tau_1 + \bm\sigma\bm h$. Here $\hat\tau_i$ and $\hat\sigma_i$ ($i=0,1,2,3$) are Pauli matrices in Nambu and spin space, $h$ is the exchange field. 
The (anti)commutator, convolution
product and differential superoperator
in Eq.~\eqref{Eq:UsadelGen} are
\begin{align}
\{\hat\tau_3\partial_\tau, \check g\}_\tau &= \hat\tau_3\partial_{\tau_1} \check g(\tau_1,\tau_2)+\partial_{\tau_2} \check g(\tau_1,\tau_2)\hat\tau_3,\\
(f\circ g)(\tau_1,\tau_2) &= \int_0^\beta d{\tau_3} f(\tau_1,\tau_3)g(\tau_3,\tau_2),\\
\hat \partial_{\bm r} &= \partial_{\bm r} - i e [\hat\tau_3 A(t), \cdot\; ].
\end{align}

Equation \eqref{Eq:UsadelGen} is complemented by the normalization condition $\check g\circ \check{g}=1$.   
The bulk charge current is given by
   \begin{align} \label{Eq:Current2}
 & {\bm j} (t) = i \frac{\sigma_n}{8e}  {\rm Tr}[\hat\tau_3 \check g \circ \hat\partial_{\bm r} \check g ] ,
 \end{align}
  where $\sigma_n=e^2\nu_0 D$ is the normal metal conductivity and  $\nu_0$ is the density of states at the Fermi level. 
 
 {\it Direct coupling to the vector potential.}
   In the dirty limit we can find corrections from the Usadel equation.
   In the frequency domain $\hat g_{AA}(\tau,\tau^\prime) = e^{i(\omega_+ \tau - \omega_- \tau^\prime)}  D\left(\frac{eA_\Omega}{c}\right)^2 \hat g_{AA} $
  that yields
  \begin{align} 
 & s_+\hat{g}_0(\omega_+) \hat g_{AA} - 
  s_- \hat g_{AA} \hat g_0(\omega_-) =
  \\ \nonumber
 &  [ \hat g_0 (\omega_+) \tau_3 \hat g_0 (\omega)\tau_3  - 
  \tau_3 \hat g_0 (\omega)\tau_3\hat g_0 (\omega_-)  ],
   \end{align}
 where we denote again 
 $\omega_\pm = \omega \pm \Omega$.  
 The solution of this equation can be written as 
  \begin{align}
  & g_{AA} =  \frac{\hat\tau_3\hat g_0 (\omega)\hat\tau_3 - \hat g_0 (\omega_+)\hat\tau_3 
  \hat g_0 (\omega) \hat\tau_3 \hat g_0(\omega_-)}
  {s_{+} + s_{-}}\label{eq:gAA}.
   \end{align}  

  {\it Contribution of the Higgs mode.}  
 Besides the corrections to the GF induced directly by the electromagnetic field 
   we need to take into account the time-dependent order parameter amplitude which drives the system out of 
   equilibrium.  
   
    The first order correction due to the external gap perturbation $\Delta_{2\Omega} e^{2i\Omega t}$ is 
       \begin{align}
   \hat g_{\Delta}   = 
   \Delta_{2\Omega}
   \frac{ \hat g_0(\omega_+)\hat\tau_2 \hat g_0(\omega_-)-\hat\tau_2}
   {s_+ + s_- },
   \end{align}  
   where $s_{\pm} = s(\omega_{\pm})$ and 
   $\omega_\pm = \omega \pm \Omega$. 
   
     The order parameter amplitude in turn is perturbed by the electric field which induces the corrections to the GF $\hat g_{AA}$
  found above in Eq.~\eqref{eq:gAA}. First we calculate the direct coupling of the order parameter amplitude to the external field described by the 
  response function
     \begin{align}
     F_\Delta (2\Omega) =  - \lambda T \sum_\omega {\rm Tr} [ \hat\tau_2 \hat g_{AA} ] 
     \end{align}
     where we introduce the dimensionless pairing constant $\lambda$.
     To find the total order parameter perturbation it is crucial to 
     renormalise the response by the corrections described by the polarization operator $\tilde{\Delta} = F_\Delta /(1-\Pi)$
      \begin{align}
      \Pi(\Omega) =  
       1+2\pi\lambda T \sum_\omega \frac{\Delta^2 + \Omega^2}{s(\omega)(\omega^2 - \Omega^2)} .
      \end{align} 
   Assuming that order parameter has frequency $2\Omega$ we  can find the corresponding perturbation of the GF 
   \begin{align}
  \hat g_{\Delta} (\tau,\tau^\prime)=   \sum_\omega \hat g_{\Delta} 
  e^{i\omega_+\tau - i \omega_-\tau^\prime}
   \end{align}    
  The first order correction to the external gap perturbation $F_\Delta$ is given by 
       \begin{align}
   \hat g_{\Delta} =  \frac{F_\Delta }{1-\Pi(2\Omega)} 
   \frac{ \hat g_0(\omega_+)\hat\tau_2 \hat g_0(\omega_-)-\hat\tau_2}
   {s_+ + s_- }.
   \end{align}

\subsection{Symmetries of the solutions and the current as functions of macroscopic fields}

 {\it Particle-hole symmetry.}
 The time-dependent quasiclassical Eq.~\eqref{Eq:UsadelGen} has the particle-hole symmetry which yields the general relation for the momentum-averaged GF
\begin{align} \label{Eq:gSymmSM}
     \check g(\bm A, \bm h, \Delta) = - \hat\tau_1 \check g(- \bm A, \bm h, \Delta^*) \hat\tau_1. 
\end{align}

{\it Artificial BdG  symmetry.}
 The weak-coupling theory of superconductivity based on the BdG equation yields an additional symmetry of the spectrum and GF  determined by the transformation 
 \begin{equation}
 \label{Eq:Transformation}
   \check g(\bm A, \bm h, \Delta) = - \hat\sigma_2 \hat\tau_1 \check g^\intercal(\bm A, \bm h, \Delta) \hat\tau_1 \hat\sigma_2. 
 \end{equation}
 Here we introduce the generalized transposition operator $\intercal$ which interchanges all indices including time.
 This transformation leaves the Usadel equation invariant. Besides that it also leaves invariant both the bulk and tunnel currents and therefore it does not provide any additional constraints for the non-linear responses.

\bibliography{refs}

\begin{thebibliography}{98}%
\makeatletter
\providecommand \@ifxundefined [1]{%
 \@ifx{#1\undefined}
}%
\providecommand \@ifnum [1]{%
 \ifnum #1\expandafter \@firstoftwo
 \else \expandafter \@secondoftwo
 \fi
}%
\providecommand \@ifx [1]{%
 \ifx #1\expandafter \@firstoftwo
 \else \expandafter \@secondoftwo
 \fi
}%
\providecommand \natexlab [1]{#1}%
\providecommand \enquote  [1]{``#1''}%
\providecommand \bibnamefont  [1]{#1}%
\providecommand \bibfnamefont [1]{#1}%
\providecommand \citenamefont [1]{#1}%
\providecommand \href@noop [0]{\@secondoftwo}%
\providecommand \href [0]{\begingroup \@sanitize@url \@href}%
\providecommand \@href[1]{\@@startlink{#1}\@@href}%
\providecommand \@@href[1]{\endgroup#1\@@endlink}%
\providecommand \@sanitize@url [0]{\catcode `\\12\catcode `\$12\catcode
  `\&12\catcode `\#12\catcode `\^12\catcode `\_12\catcode `\%12\relax}%
\providecommand \@@startlink[1]{}%
\providecommand \@@endlink[0]{}%
\providecommand \url  [0]{\begingroup\@sanitize@url \@url }%
\providecommand \@url [1]{\endgroup\@href {#1}{\urlprefix }}%
\providecommand \urlprefix  [0]{URL }%
\providecommand \Eprint [0]{\href }%
\providecommand \doibase [0]{http://dx.doi.org/}%
\providecommand \selectlanguage [0]{\@gobble}%
\providecommand \bibinfo  [0]{\@secondoftwo}%
\providecommand \bibfield  [0]{\@secondoftwo}%
\providecommand \translation [1]{[#1]}%
\providecommand \BibitemOpen [0]{}%
\providecommand \bibitemStop [0]{}%
\providecommand \bibitemNoStop [0]{.\EOS\space}%
\providecommand \EOS [0]{\spacefactor3000\relax}%
\providecommand \BibitemShut  [1]{\csname bibitem#1\endcsname}%
\let\auto@bib@innerbib\@empty
\bibitem [{\citenamefont {Varma}(2002)}]{Varma2002}%
  \BibitemOpen
  \bibfield  {author} {\bibinfo {author} {\bibfnamefont {C.~M.}\ \bibnamefont
  {Varma}},\ }\href {\doibase 10.1023/A:1013890507658} {\bibfield  {journal}
  {\bibinfo  {journal} {J. Low Temp. Phys.}\ }\textbf {\bibinfo {volume}
  {126}},\ \bibinfo {pages} {901} (\bibinfo {year} {2002})}\BibitemShut
  {NoStop}%
\bibitem [{\citenamefont {Podolsky}\ \emph {et~al.}(2011)\citenamefont
  {Podolsky}, \citenamefont {Auerbach},\ and\ \citenamefont
  {Arovas}}]{Podolsky2011}%
  \BibitemOpen
  \bibfield  {author} {\bibinfo {author} {\bibfnamefont {D.}~\bibnamefont
  {Podolsky}}, \bibinfo {author} {\bibfnamefont {A.}~\bibnamefont {Auerbach}},
  \ and\ \bibinfo {author} {\bibfnamefont {D.~P.}\ \bibnamefont {Arovas}},\
  }\href@noop {} {\bibfield  {journal} {\bibinfo  {journal} {Phys. Rev. B}\
  }\textbf {\bibinfo {volume} {84}},\ \bibinfo {pages} {174522} (\bibinfo
  {year} {2011})}\BibitemShut {NoStop}%
\bibitem [{\citenamefont {Pashkin}\ and\ \citenamefont
  {Leitenstorfer}(2014)}]{Pashkin2014}%
  \BibitemOpen
  \bibfield  {author} {\bibinfo {author} {\bibfnamefont {A.}~\bibnamefont
  {Pashkin}}\ and\ \bibinfo {author} {\bibfnamefont {A.}~\bibnamefont
  {Leitenstorfer}},\ }\href {\doibase 10.1126/science.1257302} {\bibfield
  {journal} {\bibinfo  {journal} {Science}\ }\textbf {\bibinfo {volume}
  {345}},\ \bibinfo {pages} {1121} (\bibinfo {year} {2014})}\BibitemShut
  {NoStop}%
\bibitem [{\citenamefont {Volovik}\ and\ \citenamefont
  {Zubkov}(2014)}]{Volovik2014}%
  \BibitemOpen
  \bibfield  {author} {\bibinfo {author} {\bibfnamefont {G.~E.}\ \bibnamefont
  {Volovik}}\ and\ \bibinfo {author} {\bibfnamefont {M.~A.}\ \bibnamefont
  {Zubkov}},\ }\href {\doibase 10.1007/s10909-013-0905-7} {\bibfield  {journal}
  {\bibinfo  {journal} {J. Low Temp. Phys.}\ }\textbf {\bibinfo {volume}
  {175}},\ \bibinfo {pages} {486} (\bibinfo {year} {2014})}\BibitemShut
  {NoStop}%
\bibitem [{\citenamefont {Pekker}\ and\ \citenamefont
  {Varma}(2015)}]{Varma2015}%
  \BibitemOpen
  \bibfield  {author} {\bibinfo {author} {\bibfnamefont {D.}~\bibnamefont
  {Pekker}}\ and\ \bibinfo {author} {\bibfnamefont {C.}~\bibnamefont {Varma}},\
  }\href {\doibase 10.1146/annurev-conmatphys-031214-014350} {\bibfield
  {journal} {\bibinfo  {journal} {Annu. Rev. Condens. Matter Phys.}\ }\textbf
  {\bibinfo {volume} {6}},\ \bibinfo {pages} {269} (\bibinfo {year}
  {2015})}\BibitemShut {NoStop}%
\bibitem [{\citenamefont {Higgs}(1964)}]{Higgs1964}%
  \BibitemOpen
  \bibfield  {author} {\bibinfo {author} {\bibfnamefont {P.~W.}\ \bibnamefont
  {Higgs}},\ }\href {\doibase 10.1103/PhysRevLett.13.508} {\bibfield  {journal}
  {\bibinfo  {journal} {Phys. Rev. Lett.}\ }\textbf {\bibinfo {volume} {13}},\
  \bibinfo {pages} {508} (\bibinfo {year} {1964})}\BibitemShut {NoStop}%
\bibitem [{\citenamefont {Gr\"uner}(1988)}]{RevModPhys.60.1129}%
  \BibitemOpen
  \bibfield  {author} {\bibinfo {author} {\bibfnamefont {G.}~\bibnamefont
  {Gr\"uner}},\ }\href {\doibase 10.1103/RevModPhys.60.1129} {\bibfield
  {journal} {\bibinfo  {journal} {Rev. Mod. Phys.}\ }\textbf {\bibinfo {volume}
  {60}},\ \bibinfo {pages} {1129} (\bibinfo {year} {1988})}\BibitemShut
  {NoStop}%
\bibitem [{\citenamefont {Paulson}\ \emph {et~al.}(1973)\citenamefont
  {Paulson}, \citenamefont {Johnson},\ and\ \citenamefont
  {Wheatley}}]{PhysRevLett.30.829}%
  \BibitemOpen
  \bibfield  {author} {\bibinfo {author} {\bibfnamefont {D.~N.}\ \bibnamefont
  {Paulson}}, \bibinfo {author} {\bibfnamefont {R.~T.}\ \bibnamefont
  {Johnson}}, \ and\ \bibinfo {author} {\bibfnamefont {J.~C.}\ \bibnamefont
  {Wheatley}},\ }\href {\doibase 10.1103/PhysRevLett.30.829} {\bibfield
  {journal} {\bibinfo  {journal} {Phys. Rev. Lett.}\ }\textbf {\bibinfo
  {volume} {30}},\ \bibinfo {pages} {829} (\bibinfo {year} {1973})}\BibitemShut
  {NoStop}%
\bibitem [{\citenamefont {Lawson}\ \emph {et~al.}(1973)\citenamefont {Lawson},
  \citenamefont {Gully}, \citenamefont {Goldstein}, \citenamefont
  {Richardson},\ and\ \citenamefont {Lee}}]{PhysRevLett.30.541}%
  \BibitemOpen
  \bibfield  {author} {\bibinfo {author} {\bibfnamefont {D.~T.}\ \bibnamefont
  {Lawson}}, \bibinfo {author} {\bibfnamefont {W.~J.}\ \bibnamefont {Gully}},
  \bibinfo {author} {\bibfnamefont {S.}~\bibnamefont {Goldstein}}, \bibinfo
  {author} {\bibfnamefont {R.~C.}\ \bibnamefont {Richardson}}, \ and\ \bibinfo
  {author} {\bibfnamefont {D.~M.}\ \bibnamefont {Lee}},\ }\href {\doibase
  10.1103/PhysRevLett.30.541} {\bibfield  {journal} {\bibinfo  {journal} {Phys.
  Rev. Lett.}\ }\textbf {\bibinfo {volume} {30}},\ \bibinfo {pages} {541}
  (\bibinfo {year} {1973})}\BibitemShut {NoStop}%
\bibitem [{\citenamefont {Zavjalov}\ \emph {et~al.}(2016)\citenamefont
  {Zavjalov}, \citenamefont {Autti}, \citenamefont {Eltsov}, \citenamefont
  {Heikkinen},\ and\ \citenamefont {Volovik}}]{Zavjalov2016}%
  \BibitemOpen
  \bibfield  {author} {\bibinfo {author} {\bibfnamefont {V.~V.}\ \bibnamefont
  {Zavjalov}}, \bibinfo {author} {\bibfnamefont {S.}~\bibnamefont {Autti}},
  \bibinfo {author} {\bibfnamefont {V.~B.}\ \bibnamefont {Eltsov}}, \bibinfo
  {author} {\bibfnamefont {P.~J.}\ \bibnamefont {Heikkinen}}, \ and\ \bibinfo
  {author} {\bibfnamefont {G.~E.}\ \bibnamefont {Volovik}},\ }\href {\doibase
  10.1038/ncomms10294} {\bibfield  {journal} {\bibinfo  {journal} {Nat. Comm.}\
  }\textbf {\bibinfo {volume} {7}},\ \bibinfo {pages} {10294} (\bibinfo {year}
  {2016})}\BibitemShut {NoStop}%
\bibitem [{\citenamefont {Bissbort}\ \emph {et~al.}(2011)\citenamefont
  {Bissbort}, \citenamefont {G\"otze}, \citenamefont {Li}, \citenamefont
  {Heinze}, \citenamefont {Krauser}, \citenamefont {Weinberg}, \citenamefont
  {Becker}, \citenamefont {Sengstock},\ and\ \citenamefont
  {Hofstetter}}]{PhysRevLett.106.205303}%
  \BibitemOpen
  \bibfield  {author} {\bibinfo {author} {\bibfnamefont {U.}~\bibnamefont
  {Bissbort}}, \bibinfo {author} {\bibfnamefont {S.}~\bibnamefont {G\"otze}},
  \bibinfo {author} {\bibfnamefont {Y.}~\bibnamefont {Li}}, \bibinfo {author}
  {\bibfnamefont {J.}~\bibnamefont {Heinze}}, \bibinfo {author} {\bibfnamefont
  {J.~S.}\ \bibnamefont {Krauser}}, \bibinfo {author} {\bibfnamefont
  {M.}~\bibnamefont {Weinberg}}, \bibinfo {author} {\bibfnamefont
  {C.}~\bibnamefont {Becker}}, \bibinfo {author} {\bibfnamefont
  {K.}~\bibnamefont {Sengstock}}, \ and\ \bibinfo {author} {\bibfnamefont
  {W.}~\bibnamefont {Hofstetter}},\ }\href {\doibase
  10.1103/PhysRevLett.106.205303} {\bibfield  {journal} {\bibinfo  {journal}
  {Phys. Rev. Lett.}\ }\textbf {\bibinfo {volume} {106}},\ \bibinfo {pages}
  {205303} (\bibinfo {year} {2011})}\BibitemShut {NoStop}%
\bibitem [{\citenamefont {Endres}\ \emph {et~al.}(2012)\citenamefont {Endres},
  \citenamefont {Fukuhara}, \citenamefont {Pekker}, \citenamefont {Cheneau},
  \citenamefont {Schau{\ss}}, \citenamefont {Gross}, \citenamefont {Demler},
  \citenamefont {Kuhr},\ and\ \citenamefont {Bloch}}]{Endres2012}%
  \BibitemOpen
  \bibfield  {author} {\bibinfo {author} {\bibfnamefont {M.}~\bibnamefont
  {Endres}}, \bibinfo {author} {\bibfnamefont {T.}~\bibnamefont {Fukuhara}},
  \bibinfo {author} {\bibfnamefont {D.}~\bibnamefont {Pekker}}, \bibinfo
  {author} {\bibfnamefont {M.}~\bibnamefont {Cheneau}}, \bibinfo {author}
  {\bibfnamefont {P.}~\bibnamefont {Schau{\ss}}}, \bibinfo {author}
  {\bibfnamefont {C.}~\bibnamefont {Gross}}, \bibinfo {author} {\bibfnamefont
  {E.}~\bibnamefont {Demler}}, \bibinfo {author} {\bibfnamefont
  {S.}~\bibnamefont {Kuhr}}, \ and\ \bibinfo {author} {\bibfnamefont
  {I.}~\bibnamefont {Bloch}},\ }\href {\doibase 10.1038/nature11255} {\bibfield
   {journal} {\bibinfo  {journal} {Nature}\ }\textbf {\bibinfo {volume}
  {487}},\ \bibinfo {pages} {454} (\bibinfo {year} {2012})}\BibitemShut
  {NoStop}%
\bibitem [{\citenamefont {Volkov}\ and\ \citenamefont
  {Kogan}(1974)}]{Volkov1973}%
  \BibitemOpen
  \bibfield  {author} {\bibinfo {author} {\bibfnamefont {A.~F.}\ \bibnamefont
  {Volkov}}\ and\ \bibinfo {author} {\bibfnamefont {S.~M.}\ \bibnamefont
  {Kogan}},\ }\href@noop {} {\bibfield  {journal} {\bibinfo  {journal} {JETP}\
  }\textbf {\bibinfo {volume} {38}},\ \bibinfo {pages} {1018} (\bibinfo {year}
  {1974})},\ \bibinfo {note} {[Zh. Eksp. Teor. Fiz. {\bf 65}, 2038
  (1973)]}\BibitemShut {NoStop}%
\bibitem [{\citenamefont {Sooryakumar}\ and\ \citenamefont
  {Klein}(1980)}]{Sooryakumar1980}%
  \BibitemOpen
  \bibfield  {author} {\bibinfo {author} {\bibfnamefont {R.}~\bibnamefont
  {Sooryakumar}}\ and\ \bibinfo {author} {\bibfnamefont {M.~V.}\ \bibnamefont
  {Klein}},\ }\href {\doibase 10.1103/PhysRevLett.45.660} {\bibfield  {journal}
  {\bibinfo  {journal} {Phys. Rev. Lett.}\ }\textbf {\bibinfo {volume} {45}},\
  \bibinfo {pages} {660} (\bibinfo {year} {1980})}\BibitemShut {NoStop}%
\bibitem [{\citenamefont {Littlewood}\ and\ \citenamefont
  {Varma}(1982{\natexlab{a}})}]{littlewood1982amplitude}%
  \BibitemOpen
  \bibfield  {author} {\bibinfo {author} {\bibfnamefont {P.~B.}\ \bibnamefont
  {Littlewood}}\ and\ \bibinfo {author} {\bibfnamefont {C.~M.}\ \bibnamefont
  {Varma}},\ }\href {\doibase 10.1103/PhysRevB.26.4883} {\bibfield  {journal}
  {\bibinfo  {journal} {Phys. Rev. B}\ }\textbf {\bibinfo {volume} {26}},\
  \bibinfo {pages} {4883} (\bibinfo {year} {1982}{\natexlab{a}})}\BibitemShut
  {NoStop}%
\bibitem [{\citenamefont {Barankov}\ \emph {et~al.}(2004)\citenamefont
  {Barankov}, \citenamefont {Levitov},\ and\ \citenamefont
  {Spivak}}]{Barankov2004}%
  \BibitemOpen
  \bibfield  {author} {\bibinfo {author} {\bibfnamefont {R.~A.}\ \bibnamefont
  {Barankov}}, \bibinfo {author} {\bibfnamefont {L.~S.}\ \bibnamefont
  {Levitov}}, \ and\ \bibinfo {author} {\bibfnamefont {B.~Z.}\ \bibnamefont
  {Spivak}},\ }\href {\doibase 10.1103/PhysRevLett.93.160401} {\bibfield
  {journal} {\bibinfo  {journal} {Phys. Rev. Lett.}\ }\textbf {\bibinfo
  {volume} {93}},\ \bibinfo {pages} {160401} (\bibinfo {year}
  {2004})}\BibitemShut {NoStop}%
\bibitem [{\citenamefont {Grasset}\ \emph {et~al.}(2018)\citenamefont
  {Grasset}, \citenamefont {Cea}, \citenamefont {Gallais}, \citenamefont
  {Cazayous}, \citenamefont {Sacuto}, \citenamefont {Cario}, \citenamefont
  {Benfatto},\ and\ \citenamefont {M{\'e}asson}}]{grasset2018higgs}%
  \BibitemOpen
  \bibfield  {author} {\bibinfo {author} {\bibfnamefont {R.}~\bibnamefont
  {Grasset}}, \bibinfo {author} {\bibfnamefont {T.}~\bibnamefont {Cea}},
  \bibinfo {author} {\bibfnamefont {Y.}~\bibnamefont {Gallais}}, \bibinfo
  {author} {\bibfnamefont {M.}~\bibnamefont {Cazayous}}, \bibinfo {author}
  {\bibfnamefont {A.}~\bibnamefont {Sacuto}}, \bibinfo {author} {\bibfnamefont
  {L.}~\bibnamefont {Cario}}, \bibinfo {author} {\bibfnamefont
  {L.}~\bibnamefont {Benfatto}}, \ and\ \bibinfo {author} {\bibfnamefont
  {M.-A.}\ \bibnamefont {M{\'e}asson}},\ }\href {\doibase
  10.1103/PhysRevB.97.094502} {\bibfield  {journal} {\bibinfo  {journal} {Phys.
  Rev. B}\ }\textbf {\bibinfo {volume} {97}},\ \bibinfo {pages} {094502}
  (\bibinfo {year} {2018})}\BibitemShut {NoStop}%
\bibitem [{\citenamefont {Matsunaga}\ \emph {et~al.}(2013)\citenamefont
  {Matsunaga}, \citenamefont {Hamada}, \citenamefont {Makise}, \citenamefont
  {Uzawa}, \citenamefont {Terai}, \citenamefont {Wang},\ and\ \citenamefont
  {Shimano}}]{Matsunda2013}%
  \BibitemOpen
  \bibfield  {author} {\bibinfo {author} {\bibfnamefont {R.}~\bibnamefont
  {Matsunaga}}, \bibinfo {author} {\bibfnamefont {Y.~I.}\ \bibnamefont
  {Hamada}}, \bibinfo {author} {\bibfnamefont {K.}~\bibnamefont {Makise}},
  \bibinfo {author} {\bibfnamefont {Y.}~\bibnamefont {Uzawa}}, \bibinfo
  {author} {\bibfnamefont {H.}~\bibnamefont {Terai}}, \bibinfo {author}
  {\bibfnamefont {Z.}~\bibnamefont {Wang}}, \ and\ \bibinfo {author}
  {\bibfnamefont {R.}~\bibnamefont {Shimano}},\ }\href {\doibase
  10.1103/PhysRevLett.111.057002} {\bibfield  {journal} {\bibinfo  {journal}
  {Phys. Rev. Lett.}\ }\textbf {\bibinfo {volume} {111}},\ \bibinfo {pages}
  {057002} (\bibinfo {year} {2013})}\BibitemShut {NoStop}%
\bibitem [{\citenamefont {Matsunaga}\ \emph {et~al.}(2014)\citenamefont
  {Matsunaga}, \citenamefont {Tsuji}, \citenamefont {Fujita}, \citenamefont
  {Sugioka}, \citenamefont {Makise}, \citenamefont {Uzawa}, \citenamefont
  {Terai}, \citenamefont {Wang}, \citenamefont {Aoki},\ and\ \citenamefont
  {Shimano}}]{Matsunaga1145}%
  \BibitemOpen
  \bibfield  {author} {\bibinfo {author} {\bibfnamefont {R.}~\bibnamefont
  {Matsunaga}}, \bibinfo {author} {\bibfnamefont {N.}~\bibnamefont {Tsuji}},
  \bibinfo {author} {\bibfnamefont {H.}~\bibnamefont {Fujita}}, \bibinfo
  {author} {\bibfnamefont {A.}~\bibnamefont {Sugioka}}, \bibinfo {author}
  {\bibfnamefont {K.}~\bibnamefont {Makise}}, \bibinfo {author} {\bibfnamefont
  {Y.}~\bibnamefont {Uzawa}}, \bibinfo {author} {\bibfnamefont
  {H.}~\bibnamefont {Terai}}, \bibinfo {author} {\bibfnamefont
  {Z.}~\bibnamefont {Wang}}, \bibinfo {author} {\bibfnamefont {H.}~\bibnamefont
  {Aoki}}, \ and\ \bibinfo {author} {\bibfnamefont {R.}~\bibnamefont
  {Shimano}},\ }\href {\doibase 10.1126/science.1254697} {\bibfield  {journal}
  {\bibinfo  {journal} {Science}\ }\textbf {\bibinfo {volume} {345}},\ \bibinfo
  {pages} {1145} (\bibinfo {year} {2014})}\BibitemShut {NoStop}%
\bibitem [{\citenamefont {Sherman}\ \emph {et~al.}(2015)\citenamefont
  {Sherman}, \citenamefont {Pracht}, \citenamefont {Gorshunov}, \citenamefont
  {Poran}, \citenamefont {Jesudasan}, \citenamefont {Chand}, \citenamefont
  {Raychaudhuri}, \citenamefont {Swanson}, \citenamefont {Trivedi},
  \citenamefont {Auerbach}, \citenamefont {Scheffler}, \citenamefont
  {Frydman},\ and\ \citenamefont {Dressel}}]{Sherman2015}%
  \BibitemOpen
  \bibfield  {author} {\bibinfo {author} {\bibfnamefont {D.}~\bibnamefont
  {Sherman}}, \bibinfo {author} {\bibfnamefont {U.~S.}\ \bibnamefont {Pracht}},
  \bibinfo {author} {\bibfnamefont {B.}~\bibnamefont {Gorshunov}}, \bibinfo
  {author} {\bibfnamefont {S.}~\bibnamefont {Poran}}, \bibinfo {author}
  {\bibfnamefont {J.}~\bibnamefont {Jesudasan}}, \bibinfo {author}
  {\bibfnamefont {M.}~\bibnamefont {Chand}}, \bibinfo {author} {\bibfnamefont
  {P.}~\bibnamefont {Raychaudhuri}}, \bibinfo {author} {\bibfnamefont
  {M.}~\bibnamefont {Swanson}}, \bibinfo {author} {\bibfnamefont
  {N.}~\bibnamefont {Trivedi}}, \bibinfo {author} {\bibfnamefont
  {A.}~\bibnamefont {Auerbach}}, \bibinfo {author} {\bibfnamefont
  {M.}~\bibnamefont {Scheffler}}, \bibinfo {author} {\bibfnamefont
  {A.}~\bibnamefont {Frydman}}, \ and\ \bibinfo {author} {\bibfnamefont
  {M.}~\bibnamefont {Dressel}},\ }\href {https://doi.org/10.1038/nphys3227}
  {\bibfield  {journal} {\bibinfo  {journal} {Nature Physics}\ }\textbf
  {\bibinfo {volume} {11}},\ \bibinfo {pages} {188} (\bibinfo {year}
  {2015})}\BibitemShut {NoStop}%
\bibitem [{\citenamefont {Tsuji}\ and\ \citenamefont
  {Aoki}(2015{\natexlab{a}})}]{tsuji2015theory}%
  \BibitemOpen
  \bibfield  {author} {\bibinfo {author} {\bibfnamefont {N.}~\bibnamefont
  {Tsuji}}\ and\ \bibinfo {author} {\bibfnamefont {H.}~\bibnamefont {Aoki}},\
  }\href {\doibase 10.1103/PhysRevB.92.064508} {\bibfield  {journal} {\bibinfo
  {journal} {Phys. Rev. B}\ }\textbf {\bibinfo {volume} {92}},\ \bibinfo
  {pages} {064508} (\bibinfo {year} {2015}{\natexlab{a}})}\BibitemShut
  {NoStop}%
\bibitem [{\citenamefont {Matsunaga}\ \emph {et~al.}(2017)\citenamefont
  {Matsunaga}, \citenamefont {Tsuji}, \citenamefont {Makise}, \citenamefont
  {Terai}, \citenamefont {Aoki},\ and\ \citenamefont
  {Shimano}}]{PhysRevB.96.020505}%
  \BibitemOpen
  \bibfield  {author} {\bibinfo {author} {\bibfnamefont {R.}~\bibnamefont
  {Matsunaga}}, \bibinfo {author} {\bibfnamefont {N.}~\bibnamefont {Tsuji}},
  \bibinfo {author} {\bibfnamefont {K.}~\bibnamefont {Makise}}, \bibinfo
  {author} {\bibfnamefont {H.}~\bibnamefont {Terai}}, \bibinfo {author}
  {\bibfnamefont {H.}~\bibnamefont {Aoki}}, \ and\ \bibinfo {author}
  {\bibfnamefont {R.}~\bibnamefont {Shimano}},\ }\href {\doibase
  10.1103/PhysRevB.96.020505} {\bibfield  {journal} {\bibinfo  {journal} {Phys.
  Rev. B}\ }\textbf {\bibinfo {volume} {96}},\ \bibinfo {pages} {020505}
  (\bibinfo {year} {2017})}\BibitemShut {NoStop}%
\bibitem [{\citenamefont {Katsumi}\ \emph {et~al.}(2018)\citenamefont
  {Katsumi}, \citenamefont {Tsuji}, \citenamefont {Hamada}, \citenamefont
  {Matsunaga}, \citenamefont {Schneeloch}, \citenamefont {Zhong}, \citenamefont
  {Gu}, \citenamefont {Aoki}, \citenamefont {Gallais},\ and\ \citenamefont
  {Shimano}}]{PhysRevLett.120.117001}%
  \BibitemOpen
  \bibfield  {author} {\bibinfo {author} {\bibfnamefont {K.}~\bibnamefont
  {Katsumi}}, \bibinfo {author} {\bibfnamefont {N.}~\bibnamefont {Tsuji}},
  \bibinfo {author} {\bibfnamefont {Y.~I.}\ \bibnamefont {Hamada}}, \bibinfo
  {author} {\bibfnamefont {R.}~\bibnamefont {Matsunaga}}, \bibinfo {author}
  {\bibfnamefont {J.}~\bibnamefont {Schneeloch}}, \bibinfo {author}
  {\bibfnamefont {R.~D.}\ \bibnamefont {Zhong}}, \bibinfo {author}
  {\bibfnamefont {G.~D.}\ \bibnamefont {Gu}}, \bibinfo {author} {\bibfnamefont
  {H.}~\bibnamefont {Aoki}}, \bibinfo {author} {\bibfnamefont {Y.}~\bibnamefont
  {Gallais}}, \ and\ \bibinfo {author} {\bibfnamefont {R.}~\bibnamefont
  {Shimano}},\ }\href {\doibase 10.1103/PhysRevLett.120.117001} {\bibfield
  {journal} {\bibinfo  {journal} {Phys. Rev. Lett.}\ }\textbf {\bibinfo
  {volume} {120}},\ \bibinfo {pages} {117001} (\bibinfo {year}
  {2018})}\BibitemShut {NoStop}%
\bibitem [{\citenamefont {Nakamura}\ \emph {et~al.}(2018)\citenamefont
  {Nakamura}, \citenamefont {Iida}, \citenamefont {Murotani}, \citenamefont
  {Matsunaga}, \citenamefont {Terai},\ and\ \citenamefont
  {Shimano}}]{1809.10335}%
  \BibitemOpen
  \bibfield  {author} {\bibinfo {author} {\bibfnamefont {S.}~\bibnamefont
  {Nakamura}}, \bibinfo {author} {\bibfnamefont {Y.}~\bibnamefont {Iida}},
  \bibinfo {author} {\bibfnamefont {Y.}~\bibnamefont {Murotani}}, \bibinfo
  {author} {\bibfnamefont {R.}~\bibnamefont {Matsunaga}}, \bibinfo {author}
  {\bibfnamefont {H.}~\bibnamefont {Terai}}, \ and\ \bibinfo {author}
  {\bibfnamefont {R.}~\bibnamefont {Shimano}},\ }\href@noop {} {\enquote
  {\bibinfo {title} {Infrared activation of higgs mode by supercurrent
  injection in a superconductor nbn},}\ } (\bibinfo {year} {2018}),\ \Eprint
  {http://arxiv.org/abs/arXiv:1809.10335} {arXiv:1809.10335} \BibitemShut
  {NoStop}%
\bibitem [{\citenamefont {Uematsu}\ \emph {et~al.}(2018)\citenamefont
  {Uematsu}, \citenamefont {Mizushima}, \citenamefont {Tsuruta}, \citenamefont
  {Fujimoto},\ and\ \citenamefont {Sauls}}]{1809.06989}%
  \BibitemOpen
  \bibfield  {author} {\bibinfo {author} {\bibfnamefont {H.}~\bibnamefont
  {Uematsu}}, \bibinfo {author} {\bibfnamefont {T.}~\bibnamefont {Mizushima}},
  \bibinfo {author} {\bibfnamefont {A.}~\bibnamefont {Tsuruta}}, \bibinfo
  {author} {\bibfnamefont {S.}~\bibnamefont {Fujimoto}}, \ and\ \bibinfo
  {author} {\bibfnamefont {J.~A.}\ \bibnamefont {Sauls}},\ }\href@noop {}
  {\enquote {\bibinfo {title} {Chiral higgs mode in nematic superconductors},}\
  } (\bibinfo {year} {2018}),\ \Eprint {http://arxiv.org/abs/arXiv:1809.06989}
  {arXiv:1809.06989} \BibitemShut {NoStop}%
\bibitem [{\citenamefont {Silaev}(2019{\natexlab{a}})}]{silaev2019nonlinear}%
  \BibitemOpen
  \bibfield  {author} {\bibinfo {author} {\bibfnamefont {M.}~\bibnamefont
  {Silaev}},\ }\href {\doibase 10.1103/PhysRevB.99.224511} {\bibfield
  {journal} {\bibinfo  {journal} {Phys. Rev. B}\ }\textbf {\bibinfo {volume}
  {99}},\ \bibinfo {pages} {224511} (\bibinfo {year}
  {2019}{\natexlab{a}})}\BibitemShut {NoStop}%
\bibitem [{\citenamefont {Littlewood}\ and\ \citenamefont
  {Varma}(1982{\natexlab{b}})}]{PhysRevB.26.4883}%
  \BibitemOpen
  \bibfield  {author} {\bibinfo {author} {\bibfnamefont {P.~B.}\ \bibnamefont
  {Littlewood}}\ and\ \bibinfo {author} {\bibfnamefont {C.~M.}\ \bibnamefont
  {Varma}},\ }\href {\doibase 10.1103/PhysRevB.26.4883} {\bibfield  {journal}
  {\bibinfo  {journal} {Phys. Rev. B}\ }\textbf {\bibinfo {volume} {26}},\
  \bibinfo {pages} {4883} (\bibinfo {year} {1982}{\natexlab{b}})}\BibitemShut
  {NoStop}%
\bibitem [{\citenamefont {Grasset}\ \emph {et~al.}(2019)\citenamefont
  {Grasset}, \citenamefont {Gallais}, \citenamefont {Sacuto}, \citenamefont
  {Cazayous}, \citenamefont {Ma\~nas Valero}, \citenamefont {Coronado},\ and\
  \citenamefont {M\'easson}}]{PhysRevLett.122.127001}%
  \BibitemOpen
  \bibfield  {author} {\bibinfo {author} {\bibfnamefont {R.}~\bibnamefont
  {Grasset}}, \bibinfo {author} {\bibfnamefont {Y.}~\bibnamefont {Gallais}},
  \bibinfo {author} {\bibfnamefont {A.}~\bibnamefont {Sacuto}}, \bibinfo
  {author} {\bibfnamefont {M.}~\bibnamefont {Cazayous}}, \bibinfo {author}
  {\bibfnamefont {S.}~\bibnamefont {Ma\~nas Valero}}, \bibinfo {author}
  {\bibfnamefont {E.}~\bibnamefont {Coronado}}, \ and\ \bibinfo {author}
  {\bibfnamefont {M.-A.}\ \bibnamefont {M\'easson}},\ }\href {\doibase
  10.1103/PhysRevLett.122.127001} {\bibfield  {journal} {\bibinfo  {journal}
  {Phys. Rev. Lett.}\ }\textbf {\bibinfo {volume} {122}},\ \bibinfo {pages}
  {127001} (\bibinfo {year} {2019})}\BibitemShut {NoStop}%
\bibitem [{\citenamefont {Beck}\ \emph {et~al.}(2011)\citenamefont {Beck},
  \citenamefont {Klammer}, \citenamefont {Lang}, \citenamefont {Leiderer},
  \citenamefont {Kabanov}, \citenamefont {Gol'tsman},\ and\ \citenamefont
  {Demsar}}]{PhysRevLett.107.177007}%
  \BibitemOpen
  \bibfield  {author} {\bibinfo {author} {\bibfnamefont {M.}~\bibnamefont
  {Beck}}, \bibinfo {author} {\bibfnamefont {M.}~\bibnamefont {Klammer}},
  \bibinfo {author} {\bibfnamefont {S.}~\bibnamefont {Lang}}, \bibinfo {author}
  {\bibfnamefont {P.}~\bibnamefont {Leiderer}}, \bibinfo {author}
  {\bibfnamefont {V.~V.}\ \bibnamefont {Kabanov}}, \bibinfo {author}
  {\bibfnamefont {G.~N.}\ \bibnamefont {Gol'tsman}}, \ and\ \bibinfo {author}
  {\bibfnamefont {J.}~\bibnamefont {Demsar}},\ }\href {\doibase
  10.1103/PhysRevLett.107.177007} {\bibfield  {journal} {\bibinfo  {journal}
  {Phys. Rev. Lett.}\ }\textbf {\bibinfo {volume} {107}},\ \bibinfo {pages}
  {177007} (\bibinfo {year} {2011})}\BibitemShut {NoStop}%
\bibitem [{\citenamefont {Beck}\ \emph {et~al.}(2013)\citenamefont {Beck},
  \citenamefont {Rousseau}, \citenamefont {Klammer}, \citenamefont {Leiderer},
  \citenamefont {Mittendorff}, \citenamefont {Winnerl}, \citenamefont {Helm},
  \citenamefont {Gol'tsman},\ and\ \citenamefont {Demsar}}]{beck2013transient}%
  \BibitemOpen
  \bibfield  {author} {\bibinfo {author} {\bibfnamefont {M.}~\bibnamefont
  {Beck}}, \bibinfo {author} {\bibfnamefont {I.}~\bibnamefont {Rousseau}},
  \bibinfo {author} {\bibfnamefont {M.}~\bibnamefont {Klammer}}, \bibinfo
  {author} {\bibfnamefont {P.}~\bibnamefont {Leiderer}}, \bibinfo {author}
  {\bibfnamefont {M.}~\bibnamefont {Mittendorff}}, \bibinfo {author}
  {\bibfnamefont {S.}~\bibnamefont {Winnerl}}, \bibinfo {author} {\bibfnamefont
  {M.}~\bibnamefont {Helm}}, \bibinfo {author} {\bibfnamefont {G.~N.}\
  \bibnamefont {Gol'tsman}}, \ and\ \bibinfo {author} {\bibfnamefont
  {J.}~\bibnamefont {Demsar}},\ }\href {\doibase
  10.1103/PhysRevLett.110.267003} {\bibfield  {journal} {\bibinfo  {journal}
  {Phys. Rev. Lett.}\ }\textbf {\bibinfo {volume} {110}},\ \bibinfo {pages}
  {267003} (\bibinfo {year} {2013})}\BibitemShut {NoStop}%
\bibitem [{\citenamefont {Matsunaga}\ and\ \citenamefont
  {Shimano}(2012)}]{Matsunaga2012}%
  \BibitemOpen
  \bibfield  {author} {\bibinfo {author} {\bibfnamefont {R.}~\bibnamefont
  {Matsunaga}}\ and\ \bibinfo {author} {\bibfnamefont {R.}~\bibnamefont
  {Shimano}},\ }\href {\doibase 10.1103/PhysRevLett.109.187002} {\bibfield
  {journal} {\bibinfo  {journal} {Phys. Rev. Lett.}\ }\textbf {\bibinfo
  {volume} {109}},\ \bibinfo {pages} {187002} (\bibinfo {year}
  {2012})}\BibitemShut {NoStop}%
\bibitem [{\citenamefont {Giorgianni}\ \emph {et~al.}(2019)\citenamefont
  {Giorgianni}, \citenamefont {Cea}, \citenamefont {Vicario}, \citenamefont
  {Hauri}, \citenamefont {Withanage}, \citenamefont {Xi},\ and\ \citenamefont
  {Benfatto}}]{Giorgianni2019}%
  \BibitemOpen
  \bibfield  {author} {\bibinfo {author} {\bibfnamefont {F.}~\bibnamefont
  {Giorgianni}}, \bibinfo {author} {\bibfnamefont {T.}~\bibnamefont {Cea}},
  \bibinfo {author} {\bibfnamefont {C.}~\bibnamefont {Vicario}}, \bibinfo
  {author} {\bibfnamefont {C.~P.}\ \bibnamefont {Hauri}}, \bibinfo {author}
  {\bibfnamefont {W.~K.}\ \bibnamefont {Withanage}}, \bibinfo {author}
  {\bibfnamefont {X.}~\bibnamefont {Xi}}, \ and\ \bibinfo {author}
  {\bibfnamefont {L.}~\bibnamefont {Benfatto}},\ }\href {\doibase
  10.1038/s41567-018-0385-4} {\bibfield  {journal} {\bibinfo  {journal} {Nat.
  Phys.}\ } (\bibinfo {year} {2019}),\ 10.1038/s41567-018-0385-4}\BibitemShut
  {NoStop}%
\bibitem [{\citenamefont {Kulik}\ \emph {et~al.}(1981)\citenamefont {Kulik},
  \citenamefont {Entin-Wohlman},\ and\ \citenamefont {Orbach}}]{Kulik1981}%
  \BibitemOpen
  \bibfield  {author} {\bibinfo {author} {\bibfnamefont {I.~O.}\ \bibnamefont
  {Kulik}}, \bibinfo {author} {\bibfnamefont {O.}~\bibnamefont
  {Entin-Wohlman}}, \ and\ \bibinfo {author} {\bibfnamefont {R.}~\bibnamefont
  {Orbach}},\ }\href {\doibase 10.1007/BF00115617} {\bibfield  {journal}
  {\bibinfo  {journal} {J. Low Temp. Phys.}\ }\textbf {\bibinfo {volume}
  {43}},\ \bibinfo {pages} {591} (\bibinfo {year} {1981})}\BibitemShut
  {NoStop}%
\bibitem [{\citenamefont {Barankov}\ and\ \citenamefont
  {Levitov}(2006)}]{Barankov2006}%
  \BibitemOpen
  \bibfield  {author} {\bibinfo {author} {\bibfnamefont {R.~A.}\ \bibnamefont
  {Barankov}}\ and\ \bibinfo {author} {\bibfnamefont {L.~S.}\ \bibnamefont
  {Levitov}},\ }\href {\doibase 10.1103/PhysRevLett.96.230403} {\bibfield
  {journal} {\bibinfo  {journal} {Phys. Rev. Lett.}\ }\textbf {\bibinfo
  {volume} {96}},\ \bibinfo {pages} {230403} (\bibinfo {year}
  {2006})}\BibitemShut {NoStop}%
\bibitem [{\citenamefont {Bergeret}\ \emph {et~al.}(2018)\citenamefont
  {Bergeret}, \citenamefont {Silaev}, \citenamefont {Virtanen},\ and\
  \citenamefont {Heikkil\"a}}]{RevModPhys.90.041001}%
  \BibitemOpen
  \bibfield  {author} {\bibinfo {author} {\bibfnamefont {F.~S.}\ \bibnamefont
  {Bergeret}}, \bibinfo {author} {\bibfnamefont {M.}~\bibnamefont {Silaev}},
  \bibinfo {author} {\bibfnamefont {P.}~\bibnamefont {Virtanen}}, \ and\
  \bibinfo {author} {\bibfnamefont {T.~T.}\ \bibnamefont {Heikkil\"a}},\ }\href
  {\doibase 10.1103/RevModPhys.90.041001} {\bibfield  {journal} {\bibinfo
  {journal} {Rev. Mod. Phys.}\ }\textbf {\bibinfo {volume} {90}},\ \bibinfo
  {pages} {041001} (\bibinfo {year} {2018})}\BibitemShut {NoStop}%
\bibitem [{\citenamefont {Beckmann}(2016)}]{Beckmann2016}%
  \BibitemOpen
  \bibfield  {author} {\bibinfo {author} {\bibfnamefont {D.}~\bibnamefont
  {Beckmann}},\ }\href {\doibase 10.1088/0953-8984/28/16/163001} {\bibfield
  {journal} {\bibinfo  {journal} {J. Phys. Condens. Matter}\ }\textbf {\bibinfo
  {volume} {28}},\ \bibinfo {pages} {163001} (\bibinfo {year}
  {2016})}\BibitemShut {NoStop}%
\bibitem [{\citenamefont {H{\"u}bler}\ \emph {et~al.}(2012)\citenamefont
  {H{\"u}bler}, \citenamefont {Wolf}, \citenamefont {Beckmann},\ and\
  \citenamefont {L{\"o}hneysen}}]{Huebler2012}%
  \BibitemOpen
  \bibfield  {author} {\bibinfo {author} {\bibfnamefont {F.}~\bibnamefont
  {H{\"u}bler}}, \bibinfo {author} {\bibfnamefont {M.~J.}\ \bibnamefont
  {Wolf}}, \bibinfo {author} {\bibfnamefont {D.}~\bibnamefont {Beckmann}}, \
  and\ \bibinfo {author} {\bibfnamefont {H.~v.}\ \bibnamefont
  {L{\"o}hneysen}},\ }\href {\doibase 10.1103/PhysRevLett.109.207001}
  {\bibfield  {journal} {\bibinfo  {journal} {Phys. Rev. Lett.}\ }\textbf
  {\bibinfo {volume} {109}},\ \bibinfo {pages} {207001} (\bibinfo {year}
  {2012})}\BibitemShut {NoStop}%
\bibitem [{\citenamefont {Wolf}\ \emph
  {et~al.}(2014{\natexlab{a}})\citenamefont {Wolf}, \citenamefont
  {S{\"u}rgers}, \citenamefont {Fischer},\ and\ \citenamefont
  {Beckmann}}]{Wolf2014}%
  \BibitemOpen
  \bibfield  {author} {\bibinfo {author} {\bibfnamefont {M.~J.}\ \bibnamefont
  {Wolf}}, \bibinfo {author} {\bibfnamefont {C.}~\bibnamefont {S{\"u}rgers}},
  \bibinfo {author} {\bibfnamefont {G.}~\bibnamefont {Fischer}}, \ and\
  \bibinfo {author} {\bibfnamefont {D.}~\bibnamefont {Beckmann}},\ }\href
  {\doibase 10.1103/PhysRevB.90.144509} {\bibfield  {journal} {\bibinfo
  {journal} {Phys. Rev. B}\ }\textbf {\bibinfo {volume} {90}},\ \bibinfo
  {pages} {144509} (\bibinfo {year} {2014}{\natexlab{a}})}\BibitemShut
  {NoStop}%
\bibitem [{\citenamefont {Rouco}\ \emph {et~al.}(2019)\citenamefont {Rouco},
  \citenamefont {Chakraborty}, \citenamefont {Aikebaier}, \citenamefont
  {Golovach}, \citenamefont {Strambini}, \citenamefont {Moodera}, \citenamefont
  {Giazotto}, \citenamefont {Heikkil{\"a}},\ and\ \citenamefont
  {Bergeret}}]{1906.09079}%
  \BibitemOpen
  \bibfield  {author} {\bibinfo {author} {\bibfnamefont {M.}~\bibnamefont
  {Rouco}}, \bibinfo {author} {\bibfnamefont {S.}~\bibnamefont {Chakraborty}},
  \bibinfo {author} {\bibfnamefont {F.}~\bibnamefont {Aikebaier}}, \bibinfo
  {author} {\bibfnamefont {V.~N.}\ \bibnamefont {Golovach}}, \bibinfo {author}
  {\bibfnamefont {E.}~\bibnamefont {Strambini}}, \bibinfo {author}
  {\bibfnamefont {J.~S.}\ \bibnamefont {Moodera}}, \bibinfo {author}
  {\bibfnamefont {F.}~\bibnamefont {Giazotto}}, \bibinfo {author}
  {\bibfnamefont {T.~T.}\ \bibnamefont {Heikkil{\"a}}}, \ and\ \bibinfo
  {author} {\bibfnamefont {F.~S.}\ \bibnamefont {Bergeret}},\ }\href@noop {}
  {\enquote {\bibinfo {title} {Charge transport through spin-polarized tunnel
  junction between two spin-split superconductors},}\ } (\bibinfo {year}
  {2019}),\ \Eprint {http://arxiv.org/abs/arXiv:1906.09079} {arXiv:1906.09079}
  \BibitemShut {NoStop}%
\bibitem [{\citenamefont {De~Simoni}\ \emph {et~al.}(2018)\citenamefont
  {De~Simoni}, \citenamefont {Strambini}, \citenamefont {Moodera},
  \citenamefont {Bergeret},\ and\ \citenamefont {Giazotto}}]{DeSimoni2018}%
  \BibitemOpen
  \bibfield  {author} {\bibinfo {author} {\bibfnamefont {G.}~\bibnamefont
  {De~Simoni}}, \bibinfo {author} {\bibfnamefont {E.}~\bibnamefont
  {Strambini}}, \bibinfo {author} {\bibfnamefont {J.~S.}\ \bibnamefont
  {Moodera}}, \bibinfo {author} {\bibfnamefont {F.~S.}\ \bibnamefont
  {Bergeret}}, \ and\ \bibinfo {author} {\bibfnamefont {F.}~\bibnamefont
  {Giazotto}},\ }\href {\doibase 10.1021/acs.nanolett.8b02723} {\bibfield
  {journal} {\bibinfo  {journal} {Nano Lett.}\ }\textbf {\bibinfo {volume}
  {18}},\ \bibinfo {pages} {6369} (\bibinfo {year} {2018})}\BibitemShut
  {NoStop}%
\bibitem [{\citenamefont {Strambini}\ \emph {et~al.}(2017)\citenamefont
  {Strambini}, \citenamefont {Golovach}, \citenamefont {De~Simoni},
  \citenamefont {Moodera}, \citenamefont {Bergeret},\ and\ \citenamefont
  {Giazotto}}]{PhysRevMaterials.1.054402}%
  \BibitemOpen
  \bibfield  {author} {\bibinfo {author} {\bibfnamefont {E.}~\bibnamefont
  {Strambini}}, \bibinfo {author} {\bibfnamefont {V.~N.}\ \bibnamefont
  {Golovach}}, \bibinfo {author} {\bibfnamefont {G.}~\bibnamefont {De~Simoni}},
  \bibinfo {author} {\bibfnamefont {J.~S.}\ \bibnamefont {Moodera}}, \bibinfo
  {author} {\bibfnamefont {F.~S.}\ \bibnamefont {Bergeret}}, \ and\ \bibinfo
  {author} {\bibfnamefont {F.}~\bibnamefont {Giazotto}},\ }\href {\doibase
  10.1103/PhysRevMaterials.1.054402} {\bibfield  {journal} {\bibinfo  {journal}
  {Phys. Rev. Materials}\ }\textbf {\bibinfo {volume} {1}},\ \bibinfo {pages}
  {054402} (\bibinfo {year} {2017})}\BibitemShut {NoStop}%
\bibitem [{\citenamefont {Kolenda}\ \emph {et~al.}(2016)\citenamefont
  {Kolenda}, \citenamefont {Wolf},\ and\ \citenamefont
  {Beckmann}}]{PhysRevLett.116.097001}%
  \BibitemOpen
  \bibfield  {author} {\bibinfo {author} {\bibfnamefont {S.}~\bibnamefont
  {Kolenda}}, \bibinfo {author} {\bibfnamefont {M.~J.}\ \bibnamefont {Wolf}}, \
  and\ \bibinfo {author} {\bibfnamefont {D.}~\bibnamefont {Beckmann}},\ }\href
  {\doibase 10.1103/PhysRevLett.116.097001} {\bibfield  {journal} {\bibinfo
  {journal} {Phys. Rev. Lett.}\ }\textbf {\bibinfo {volume} {116}},\ \bibinfo
  {pages} {097001} (\bibinfo {year} {2016})}\BibitemShut {NoStop}%
\bibitem [{\citenamefont {Wolf}\ \emph
  {et~al.}(2014{\natexlab{b}})\citenamefont {Wolf}, \citenamefont
  {S{\"u}rgers}, \citenamefont {Fischer},\ and\ \citenamefont
  {Beckmann}}]{PhysRevB.90.144509}%
  \BibitemOpen
  \bibfield  {author} {\bibinfo {author} {\bibfnamefont {M.~J.}\ \bibnamefont
  {Wolf}}, \bibinfo {author} {\bibfnamefont {C.}~\bibnamefont {S{\"u}rgers}},
  \bibinfo {author} {\bibfnamefont {G.}~\bibnamefont {Fischer}}, \ and\
  \bibinfo {author} {\bibfnamefont {D.}~\bibnamefont {Beckmann}},\ }\href
  {\doibase 10.1103/PhysRevB.90.144509} {\bibfield  {journal} {\bibinfo
  {journal} {Phys. Rev. B}\ }\textbf {\bibinfo {volume} {90}},\ \bibinfo
  {pages} {144509} (\bibinfo {year} {2014}{\natexlab{b}})}\BibitemShut
  {NoStop}%
\bibitem [{\citenamefont {Wolf}\ \emph {et~al.}(2013)\citenamefont {Wolf},
  \citenamefont {H\"ubler}, \citenamefont {Kolenda}, \citenamefont
  {v.~L\"ohneysen},\ and\ \citenamefont {Beckmann}}]{PhysRevB.87.024517}%
  \BibitemOpen
  \bibfield  {author} {\bibinfo {author} {\bibfnamefont {M.~J.}\ \bibnamefont
  {Wolf}}, \bibinfo {author} {\bibfnamefont {F.}~\bibnamefont {H\"ubler}},
  \bibinfo {author} {\bibfnamefont {S.}~\bibnamefont {Kolenda}}, \bibinfo
  {author} {\bibfnamefont {H.}~\bibnamefont {v.~L\"ohneysen}}, \ and\ \bibinfo
  {author} {\bibfnamefont {D.}~\bibnamefont {Beckmann}},\ }\href {\doibase
  10.1103/PhysRevB.87.024517} {\bibfield  {journal} {\bibinfo  {journal} {Phys.
  Rev. B}\ }\textbf {\bibinfo {volume} {87}},\ \bibinfo {pages} {024517}
  (\bibinfo {year} {2013})}\BibitemShut {NoStop}%
\bibitem [{\citenamefont {Quay}\ \emph {et~al.}(2013)\citenamefont {Quay},
  \citenamefont {Chevallier}, \citenamefont {Bena},\ and\ \citenamefont
  {Aprili}}]{Quay2013}%
  \BibitemOpen
  \bibfield  {author} {\bibinfo {author} {\bibfnamefont {C.~H.~L.}\
  \bibnamefont {Quay}}, \bibinfo {author} {\bibfnamefont {D.}~\bibnamefont
  {Chevallier}}, \bibinfo {author} {\bibfnamefont {C.}~\bibnamefont {Bena}}, \
  and\ \bibinfo {author} {\bibfnamefont {M.}~\bibnamefont {Aprili}},\ }\href
  {\doibase 10.1038/nphys2518} {\bibfield  {journal} {\bibinfo  {journal} {Nat.
  Phys.}\ }\textbf {\bibinfo {volume} {9}},\ \bibinfo {pages} {84} (\bibinfo
  {year} {2013})}\BibitemShut {NoStop}%
\bibitem [{\citenamefont {Quay}\ \emph {et~al.}(2016)\citenamefont {Quay},
  \citenamefont {Dutreix}, \citenamefont {Chevallier}, \citenamefont {Bena},\
  and\ \citenamefont {Aprili}}]{Quay2016}%
  \BibitemOpen
  \bibfield  {author} {\bibinfo {author} {\bibfnamefont {C.~H.~L.}\
  \bibnamefont {Quay}}, \bibinfo {author} {\bibfnamefont {C.}~\bibnamefont
  {Dutreix}}, \bibinfo {author} {\bibfnamefont {D.}~\bibnamefont {Chevallier}},
  \bibinfo {author} {\bibfnamefont {C.}~\bibnamefont {Bena}}, \ and\ \bibinfo
  {author} {\bibfnamefont {M.}~\bibnamefont {Aprili}},\ }\href {\doibase
  10.1103/PhysRevB.93.220501} {\bibfield  {journal} {\bibinfo  {journal} {Phys.
  Rev. B}\ }\textbf {\bibinfo {volume} {93}},\ \bibinfo {pages} {220501}
  (\bibinfo {year} {2016})}\BibitemShut {NoStop}%
\bibitem [{\citenamefont {Silaev}\ \emph
  {et~al.}(2015{\natexlab{a}})\citenamefont {Silaev}, \citenamefont {Virtanen},
  \citenamefont {Bergeret},\ and\ \citenamefont {Heikkil{\"a}}}]{Silaev2015}%
  \BibitemOpen
  \bibfield  {author} {\bibinfo {author} {\bibfnamefont {M.}~\bibnamefont
  {Silaev}}, \bibinfo {author} {\bibfnamefont {P.}~\bibnamefont {Virtanen}},
  \bibinfo {author} {\bibfnamefont {F.~S.}\ \bibnamefont {Bergeret}}, \ and\
  \bibinfo {author} {\bibfnamefont {T.~T.}\ \bibnamefont {Heikkil{\"a}}},\
  }\href {\doibase 10.1103/PhysRevLett.114.167002} {\bibfield  {journal}
  {\bibinfo  {journal} {Phys. Rev. Lett.}\ }\textbf {\bibinfo {volume} {114}},\
  \bibinfo {pages} {167002} (\bibinfo {year} {2015}{\natexlab{a}})}\BibitemShut
  {NoStop}%
\bibitem [{\citenamefont {Bobkova}\ and\ \citenamefont
  {Bobkov}(2015)}]{Bobkova2015}%
  \BibitemOpen
  \bibfield  {author} {\bibinfo {author} {\bibfnamefont {I.~V.}\ \bibnamefont
  {Bobkova}}\ and\ \bibinfo {author} {\bibfnamefont {A.~M.}\ \bibnamefont
  {Bobkov}},\ }\href {\doibase 10.1134/S0021364015020022} {\bibfield  {journal}
  {\bibinfo  {journal} {JETP Letters}\ }\textbf {\bibinfo {volume} {101}},\
  \bibinfo {pages} {118} (\bibinfo {year} {2015})}\BibitemShut {NoStop}%
\bibitem [{\citenamefont {Krishtop}\ \emph {et~al.}(2015)\citenamefont
  {Krishtop}, \citenamefont {Houzet},\ and\ \citenamefont
  {Meyer}}]{Krishtop2015}%
  \BibitemOpen
  \bibfield  {author} {\bibinfo {author} {\bibfnamefont {T.}~\bibnamefont
  {Krishtop}}, \bibinfo {author} {\bibfnamefont {M.}~\bibnamefont {Houzet}}, \
  and\ \bibinfo {author} {\bibfnamefont {J.~S.}\ \bibnamefont {Meyer}},\ }\href
  {\doibase 10.1103/PhysRevB.91.121407} {\bibfield  {journal} {\bibinfo
  {journal} {Physical Review B}\ }\textbf {\bibinfo {volume} {91}},\ \bibinfo
  {pages} {121407} (\bibinfo {year} {2015})}\BibitemShut {NoStop}%
\bibitem [{\citenamefont {Virtanen}\ \emph {et~al.}(2016)\citenamefont
  {Virtanen}, \citenamefont {Heikkil\"a},\ and\ \citenamefont
  {Bergeret}}]{PhysRevB.93.014512}%
  \BibitemOpen
  \bibfield  {author} {\bibinfo {author} {\bibfnamefont {P.}~\bibnamefont
  {Virtanen}}, \bibinfo {author} {\bibfnamefont {T.~T.}\ \bibnamefont
  {Heikkil\"a}}, \ and\ \bibinfo {author} {\bibfnamefont {F.~S.}\ \bibnamefont
  {Bergeret}},\ }\href {\doibase 10.1103/PhysRevB.93.014512} {\bibfield
  {journal} {\bibinfo  {journal} {Phys. Rev. B}\ }\textbf {\bibinfo {volume}
  {93}},\ \bibinfo {pages} {014512} (\bibinfo {year} {2016})}\BibitemShut
  {NoStop}%
\bibitem [{\citenamefont {Aikebaier}\ \emph {et~al.}(2018)\citenamefont
  {Aikebaier}, \citenamefont {Silaev},\ and\ \citenamefont
  {Heikkil{\"a}}}]{PhysRevB.98.024516}%
  \BibitemOpen
  \bibfield  {author} {\bibinfo {author} {\bibfnamefont {F.}~\bibnamefont
  {Aikebaier}}, \bibinfo {author} {\bibfnamefont {M.~A.}\ \bibnamefont
  {Silaev}}, \ and\ \bibinfo {author} {\bibfnamefont {T.~T.}\ \bibnamefont
  {Heikkil{\"a}}},\ }\href {\doibase 10.1103/PhysRevB.98.024516} {\bibfield
  {journal} {\bibinfo  {journal} {Phys. Rev. B}\ }\textbf {\bibinfo {volume}
  {98}},\ \bibinfo {pages} {024516} (\bibinfo {year} {2018})}\BibitemShut
  {NoStop}%
\bibitem [{\citenamefont {Virtanen}\ \emph {et~al.}(2018)\citenamefont
  {Virtanen}, \citenamefont {Bergeret}, \citenamefont {Strambini},
  \citenamefont {Giazotto},\ and\ \citenamefont
  {Braggio}}]{PhysRevB.98.020501}%
  \BibitemOpen
  \bibfield  {author} {\bibinfo {author} {\bibfnamefont {P.}~\bibnamefont
  {Virtanen}}, \bibinfo {author} {\bibfnamefont {F.~S.}\ \bibnamefont
  {Bergeret}}, \bibinfo {author} {\bibfnamefont {E.}~\bibnamefont {Strambini}},
  \bibinfo {author} {\bibfnamefont {F.}~\bibnamefont {Giazotto}}, \ and\
  \bibinfo {author} {\bibfnamefont {A.}~\bibnamefont {Braggio}},\ }\href
  {\doibase 10.1103/PhysRevB.98.020501} {\bibfield  {journal} {\bibinfo
  {journal} {Phys. Rev. B}\ }\textbf {\bibinfo {volume} {98}},\ \bibinfo
  {pages} {020501} (\bibinfo {year} {2018})}\BibitemShut {NoStop}%
\bibitem [{\citenamefont {Bobkova}\ and\ \citenamefont
  {Bobkov}(2016)}]{PhysRevB.93.024513}%
  \BibitemOpen
  \bibfield  {author} {\bibinfo {author} {\bibfnamefont {I.~V.}\ \bibnamefont
  {Bobkova}}\ and\ \bibinfo {author} {\bibfnamefont {A.~M.}\ \bibnamefont
  {Bobkov}},\ }\href {\doibase 10.1103/PhysRevB.93.024513} {\bibfield
  {journal} {\bibinfo  {journal} {Phys. Rev. B}\ }\textbf {\bibinfo {volume}
  {93}},\ \bibinfo {pages} {024513} (\bibinfo {year} {2016})}\BibitemShut
  {NoStop}%
\bibitem [{\citenamefont {Kim}\ \emph {et~al.}(2018)\citenamefont {Kim},
  \citenamefont {Myers},\ and\ \citenamefont
  {Tserkovnyak}}]{PhysRevLett.121.187203}%
  \BibitemOpen
  \bibfield  {author} {\bibinfo {author} {\bibfnamefont {S.~K.}\ \bibnamefont
  {Kim}}, \bibinfo {author} {\bibfnamefont {R.}~\bibnamefont {Myers}}, \ and\
  \bibinfo {author} {\bibfnamefont {Y.}~\bibnamefont {Tserkovnyak}},\ }\href
  {\doibase 10.1103/PhysRevLett.121.187203} {\bibfield  {journal} {\bibinfo
  {journal} {Phys. Rev. Lett.}\ }\textbf {\bibinfo {volume} {121}},\ \bibinfo
  {pages} {187203} (\bibinfo {year} {2018})}\BibitemShut {NoStop}%
\bibitem [{\citenamefont {Vargunin}\ and\ \citenamefont
  {Silaev}(2019)}]{Vargunin2019}%
  \BibitemOpen
  \bibfield  {author} {\bibinfo {author} {\bibfnamefont {A.}~\bibnamefont
  {Vargunin}}\ and\ \bibinfo {author} {\bibfnamefont {M.}~\bibnamefont
  {Silaev}},\ }\href {\doibase 10.1038/s41598-019-42034-y} {\bibfield
  {journal} {\bibinfo  {journal} {Sci. Rep.}\ }\textbf {\bibinfo {volume}
  {9}},\ \bibinfo {pages} {5914} (\bibinfo {year} {2019})}\BibitemShut
  {NoStop}%
\bibitem [{\citenamefont {Bergeret}\ \emph {et~al.}(2001)\citenamefont
  {Bergeret}, \citenamefont {Volkov},\ and\ \citenamefont
  {Efetov}}]{PhysRevLett.86.3140}%
  \BibitemOpen
  \bibfield  {author} {\bibinfo {author} {\bibfnamefont {F.~S.}\ \bibnamefont
  {Bergeret}}, \bibinfo {author} {\bibfnamefont {A.~F.}\ \bibnamefont
  {Volkov}}, \ and\ \bibinfo {author} {\bibfnamefont {K.~B.}\ \bibnamefont
  {Efetov}},\ }\href {\doibase 10.1103/PhysRevLett.86.3140} {\bibfield
  {journal} {\bibinfo  {journal} {Phys. Rev. Lett.}\ }\textbf {\bibinfo
  {volume} {86}},\ \bibinfo {pages} {3140} (\bibinfo {year}
  {2001})}\BibitemShut {NoStop}%
\bibitem [{\citenamefont {Tokuyasu}\ \emph {et~al.}(1988)\citenamefont
  {Tokuyasu}, \citenamefont {Sauls},\ and\ \citenamefont
  {Rainer}}]{PhysRevB.38.8823}%
  \BibitemOpen
  \bibfield  {author} {\bibinfo {author} {\bibfnamefont {T.}~\bibnamefont
  {Tokuyasu}}, \bibinfo {author} {\bibfnamefont {J.~A.}\ \bibnamefont {Sauls}},
  \ and\ \bibinfo {author} {\bibfnamefont {D.}~\bibnamefont {Rainer}},\ }\href
  {\doibase 10.1103/PhysRevB.38.8823} {\bibfield  {journal} {\bibinfo
  {journal} {Phys. Rev. B}\ }\textbf {\bibinfo {volume} {38}},\ \bibinfo
  {pages} {8823} (\bibinfo {year} {1988})}\BibitemShut {NoStop}%
\bibitem [{\citenamefont {Millis}\ \emph {et~al.}(1988)\citenamefont {Millis},
  \citenamefont {Rainer},\ and\ \citenamefont {Sauls}}]{PhysRevB.38.4504}%
  \BibitemOpen
  \bibfield  {author} {\bibinfo {author} {\bibfnamefont {A.}~\bibnamefont
  {Millis}}, \bibinfo {author} {\bibfnamefont {D.}~\bibnamefont {Rainer}}, \
  and\ \bibinfo {author} {\bibfnamefont {J.~A.}\ \bibnamefont {Sauls}},\ }\href
  {\doibase 10.1103/PhysRevB.38.4504} {\bibfield  {journal} {\bibinfo
  {journal} {Phys. Rev. B}\ }\textbf {\bibinfo {volume} {38}},\ \bibinfo
  {pages} {4504} (\bibinfo {year} {1988})}\BibitemShut {NoStop}%
\bibitem [{\citenamefont {Cottet}\ \emph {et~al.}(2009)\citenamefont {Cottet},
  \citenamefont {Huertas-Hernando}, \citenamefont {Belzig},\ and\ \citenamefont
  {Nazarov}}]{PhysRevB.80.184511}%
  \BibitemOpen
  \bibfield  {author} {\bibinfo {author} {\bibfnamefont {A.}~\bibnamefont
  {Cottet}}, \bibinfo {author} {\bibfnamefont {D.}~\bibnamefont
  {Huertas-Hernando}}, \bibinfo {author} {\bibfnamefont {W.}~\bibnamefont
  {Belzig}}, \ and\ \bibinfo {author} {\bibfnamefont {Y.~V.}\ \bibnamefont
  {Nazarov}},\ }\href {\doibase 10.1103/PhysRevB.80.184511} {\bibfield
  {journal} {\bibinfo  {journal} {Phys. Rev. B}\ }\textbf {\bibinfo {volume}
  {80}},\ \bibinfo {pages} {184511} (\bibinfo {year} {2009})}\BibitemShut
  {NoStop}%
\bibitem [{\citenamefont {Eschrig}\ \emph {et~al.}(2015)\citenamefont
  {Eschrig}, \citenamefont {Cottet}, \citenamefont {Belzig},\ and\
  \citenamefont {Linder}}]{Eschrig_2015}%
  \BibitemOpen
  \bibfield  {author} {\bibinfo {author} {\bibfnamefont {M.}~\bibnamefont
  {Eschrig}}, \bibinfo {author} {\bibfnamefont {A.}~\bibnamefont {Cottet}},
  \bibinfo {author} {\bibfnamefont {W.}~\bibnamefont {Belzig}}, \ and\ \bibinfo
  {author} {\bibfnamefont {J.}~\bibnamefont {Linder}},\ }\href {\doibase
  10.1088/1367-2630/17/8/083037} {\bibfield  {journal} {\bibinfo  {journal}
  {New J. Phys.}\ }\textbf {\bibinfo {volume} {17}},\ \bibinfo {pages} {083037}
  (\bibinfo {year} {2015})}\BibitemShut {NoStop}%
\bibitem [{\citenamefont {Gorkov}\ and\ \citenamefont
  {Eliashberg}(1968)}]{Gorkov1968}%
  \BibitemOpen
  \bibfield  {author} {\bibinfo {author} {\bibfnamefont {L.~P.}\ \bibnamefont
  {Gorkov}}\ and\ \bibinfo {author} {\bibfnamefont {G.~M.}\ \bibnamefont
  {Eliashberg}},\ }\href@noop {} {\bibfield  {journal} {\bibinfo  {journal}
  {JETP}\ }\textbf {\bibinfo {volume} {27}},\ \bibinfo {pages} {328} (\bibinfo
  {year} {1968})}\BibitemShut {NoStop}%
\bibitem [{\citenamefont {Gorkov}\ and\ \citenamefont
  {Eliashberg}(1969)}]{Gorkov1969}%
  \BibitemOpen
  \bibfield  {author} {\bibinfo {author} {\bibfnamefont {L.~P.}\ \bibnamefont
  {Gorkov}}\ and\ \bibinfo {author} {\bibfnamefont {G.~M.}\ \bibnamefont
  {Eliashberg}},\ }\href@noop {} {\bibfield  {journal} {\bibinfo  {journal}
  {JETP}\ }\textbf {\bibinfo {volume} {28}},\ \bibinfo {pages} {1291} (\bibinfo
  {year} {1969})}\BibitemShut {NoStop}%
\bibitem [{\citenamefont {Dynes}\ \emph {et~al.}(1984)\citenamefont {Dynes},
  \citenamefont {Garno}, \citenamefont {Hertel},\ and\ \citenamefont
  {Orlando}}]{dynes84}%
  \BibitemOpen
  \bibfield  {author} {\bibinfo {author} {\bibfnamefont {R.~C.}\ \bibnamefont
  {Dynes}}, \bibinfo {author} {\bibfnamefont {J.~P.}\ \bibnamefont {Garno}},
  \bibinfo {author} {\bibfnamefont {G.~B.}\ \bibnamefont {Hertel}}, \ and\
  \bibinfo {author} {\bibfnamefont {T.~P.}\ \bibnamefont {Orlando}},\ }\href
  {\doibase 10.1103/PhysRevLett.53.2437} {\bibfield  {journal} {\bibinfo
  {journal} {Phys. Rev. Lett.}\ }\textbf {\bibinfo {volume} {53}},\ \bibinfo
  {pages} {2437} (\bibinfo {year} {1984})}\BibitemShut {NoStop}%
\bibitem [{Sup()}]{SupplementaryMaterial}%
  \BibitemOpen
  \href@noop {} {}\bibinfo {note} {Supplementary material file includes
  derivation of the general expression for nonstationary tunneling current,
  calculations of the nonlinear responses in terms of the Usadel equation,
  discussion of the symmetries for the solutions and currents as functions of
  the macroscopic fields.}\BibitemShut {Stop}%
\bibitem [{Note1()}]{Note1}%
  \BibitemOpen
  \bibinfo {note} {Here we exclude the trivial SHG generation which results
  from the third-order nonlinearity when both the oscillating and constant
  fields are applied}\BibitemShut {NoStop}%
\bibitem [{\citenamefont {Amato}\ and\ \citenamefont
  {McLean}(1976)}]{PhysRevLett.37.930}%
  \BibitemOpen
  \bibfield  {author} {\bibinfo {author} {\bibfnamefont {J.~C.}\ \bibnamefont
  {Amato}}\ and\ \bibinfo {author} {\bibfnamefont {W.~L.}\ \bibnamefont
  {McLean}},\ }\href {\doibase 10.1103/PhysRevLett.37.930} {\bibfield
  {journal} {\bibinfo  {journal} {Phys. Rev. Lett.}\ }\textbf {\bibinfo
  {volume} {37}},\ \bibinfo {pages} {930} (\bibinfo {year} {1976})}\BibitemShut
  {NoStop}%
\bibitem [{\citenamefont {Silaev}\ \emph {et~al.}(2017)\citenamefont {Silaev},
  \citenamefont {Tokatly},\ and\ \citenamefont {Bergeret}}]{Silaev2017}%
  \BibitemOpen
  \bibfield  {author} {\bibinfo {author} {\bibfnamefont {M.~A.}\ \bibnamefont
  {Silaev}}, \bibinfo {author} {\bibfnamefont {I.~V.}\ \bibnamefont {Tokatly}},
  \ and\ \bibinfo {author} {\bibfnamefont {F.~S.}\ \bibnamefont {Bergeret}},\
  }\href {https://link.aps.org/doi/10.1103/PhysRevB.95.184508} {\bibfield
  {journal} {\bibinfo  {journal} {Phys. Rev. B}\ }\textbf {\bibinfo {volume}
  {95}},\ \bibinfo {pages} {184508} (\bibinfo {year} {2017})}\BibitemShut
  {NoStop}%
\bibitem [{qua()}]{quasiclassicalnote}%
  \BibitemOpen
  \href@noop {} {}\bibinfo {note} {The Fermi surface symmetry present in the
  quasiclassical approximation is broken here by the spin polarization of
  tunneling.}\BibitemShut {Stop}%
\bibitem [{\citenamefont {Machon}\ \emph {et~al.}(2013)\citenamefont {Machon},
  \citenamefont {Eschrig},\ and\ \citenamefont
  {Belzig}}]{PhysRevLett.110.047002}%
  \BibitemOpen
  \bibfield  {author} {\bibinfo {author} {\bibfnamefont {P.}~\bibnamefont
  {Machon}}, \bibinfo {author} {\bibfnamefont {M.}~\bibnamefont {Eschrig}}, \
  and\ \bibinfo {author} {\bibfnamefont {W.}~\bibnamefont {Belzig}},\ }\href
  {\doibase 10.1103/PhysRevLett.110.047002} {\bibfield  {journal} {\bibinfo
  {journal} {Phys. Rev. Lett.}\ }\textbf {\bibinfo {volume} {110}},\ \bibinfo
  {pages} {047002} (\bibinfo {year} {2013})}\BibitemShut {NoStop}%
\bibitem [{\citenamefont {Ozaeta}\ \emph {et~al.}(2014)\citenamefont {Ozaeta},
  \citenamefont {Virtanen}, \citenamefont {Bergeret},\ and\ \citenamefont
  {Heikkil{\"a}}}]{ozaeta2014predicted}%
  \BibitemOpen
  \bibfield  {author} {\bibinfo {author} {\bibfnamefont {A.}~\bibnamefont
  {Ozaeta}}, \bibinfo {author} {\bibfnamefont {P.}~\bibnamefont {Virtanen}},
  \bibinfo {author} {\bibfnamefont {F.~S.}\ \bibnamefont {Bergeret}}, \ and\
  \bibinfo {author} {\bibfnamefont {T.~T.}\ \bibnamefont {Heikkil{\"a}}},\
  }\href {\doibase 10.1103/PhysRevLett.112.057001} {\bibfield  {journal}
  {\bibinfo  {journal} {Phys. Rev. Lett.}\ }\textbf {\bibinfo {volume} {112}},\
  \bibinfo {pages} {057001} (\bibinfo {year} {2014})}\BibitemShut {NoStop}%
\bibitem [{\citenamefont {Ambegaokar}\ and\ \citenamefont
  {Baratoff}(1963)}]{PhysRevLett.10.486}%
  \BibitemOpen
  \bibfield  {author} {\bibinfo {author} {\bibfnamefont {V.}~\bibnamefont
  {Ambegaokar}}\ and\ \bibinfo {author} {\bibfnamefont {A.}~\bibnamefont
  {Baratoff}},\ }\href {\doibase 10.1103/PhysRevLett.10.486} {\bibfield
  {journal} {\bibinfo  {journal} {Phys. Rev. Lett.}\ }\textbf {\bibinfo
  {volume} {10}},\ \bibinfo {pages} {486} (\bibinfo {year} {1963})}\BibitemShut
  {NoStop}%
\bibitem [{\citenamefont {Bardeen}(1962)}]{PhysRevLett.9.147}%
  \BibitemOpen
  \bibfield  {author} {\bibinfo {author} {\bibfnamefont {J.}~\bibnamefont
  {Bardeen}},\ }\href {\doibase 10.1103/PhysRevLett.9.147} {\bibfield
  {journal} {\bibinfo  {journal} {Phys. Rev. Lett.}\ }\textbf {\bibinfo
  {volume} {9}},\ \bibinfo {pages} {147} (\bibinfo {year} {1962})}\BibitemShut
  {NoStop}%
\bibitem [{\citenamefont {Eckern}\ \emph {et~al.}(1984)\citenamefont {Eckern},
  \citenamefont {Sch\"on},\ and\ \citenamefont
  {Ambegaokar}}]{PhysRevB.30.6419}%
  \BibitemOpen
  \bibfield  {author} {\bibinfo {author} {\bibfnamefont {U.}~\bibnamefont
  {Eckern}}, \bibinfo {author} {\bibfnamefont {G.}~\bibnamefont {Sch\"on}}, \
  and\ \bibinfo {author} {\bibfnamefont {V.}~\bibnamefont {Ambegaokar}},\
  }\href {\doibase 10.1103/PhysRevB.30.6419} {\bibfield  {journal} {\bibinfo
  {journal} {Phys. Rev. B}\ }\textbf {\bibinfo {volume} {30}},\ \bibinfo
  {pages} {6419} (\bibinfo {year} {1984})}\BibitemShut {NoStop}%
\bibitem [{\citenamefont {Harris}(1975)}]{PhysRevB.11.3329}%
  \BibitemOpen
  \bibfield  {author} {\bibinfo {author} {\bibfnamefont {R.~E.}\ \bibnamefont
  {Harris}},\ }\href {\doibase 10.1103/PhysRevB.11.3329} {\bibfield  {journal}
  {\bibinfo  {journal} {Phys. Rev. B}\ }\textbf {\bibinfo {volume} {11}},\
  \bibinfo {pages} {3329} (\bibinfo {year} {1975})}\BibitemShut {NoStop}%
\bibitem [{\citenamefont {Werthamer}(1966)}]{PhysRev.147.255}%
  \BibitemOpen
  \bibfield  {author} {\bibinfo {author} {\bibfnamefont {N.~R.}\ \bibnamefont
  {Werthamer}},\ }\href {\doibase 10.1103/PhysRev.147.255} {\bibfield
  {journal} {\bibinfo  {journal} {Phys. Rev.}\ }\textbf {\bibinfo {volume}
  {147}},\ \bibinfo {pages} {255} (\bibinfo {year} {1966})}\BibitemShut
  {NoStop}%
\bibitem [{\citenamefont {Bergeret}\ \emph
  {et~al.}(2012{\natexlab{a}})\citenamefont {Bergeret}, \citenamefont {Verso},\
  and\ \citenamefont {Volkov}}]{bergeret2012electronic}%
  \BibitemOpen
  \bibfield  {author} {\bibinfo {author} {\bibfnamefont {F.~S.}\ \bibnamefont
  {Bergeret}}, \bibinfo {author} {\bibfnamefont {A.}~\bibnamefont {Verso}}, \
  and\ \bibinfo {author} {\bibfnamefont {A.~F.}\ \bibnamefont {Volkov}},\
  }\href {\doibase 10.1103/PhysRevB.86.214516} {\bibfield  {journal} {\bibinfo
  {journal} {Phys. Rev. B}\ }\textbf {\bibinfo {volume} {86}},\ \bibinfo
  {pages} {214516} (\bibinfo {year} {2012}{\natexlab{a}})}\BibitemShut
  {NoStop}%
\bibitem [{\citenamefont {Bergeret}\ \emph
  {et~al.}(2012{\natexlab{b}})\citenamefont {Bergeret}, \citenamefont {Verso},\
  and\ \citenamefont {Volkov}}]{Bergeret2012a}%
  \BibitemOpen
  \bibfield  {author} {\bibinfo {author} {\bibfnamefont {F.~S.}\ \bibnamefont
  {Bergeret}}, \bibinfo {author} {\bibfnamefont {A.}~\bibnamefont {Verso}}, \
  and\ \bibinfo {author} {\bibfnamefont {A.~F.}\ \bibnamefont {Volkov}},\
  }\href {\doibase 10.1103/PhysRevB.86.060506} {\bibfield  {journal} {\bibinfo
  {journal} {Phys. Rev. B}\ }\textbf {\bibinfo {volume} {86}},\ \bibinfo
  {pages} {060506} (\bibinfo {year} {2012}{\natexlab{b}})}\BibitemShut
  {NoStop}%
\bibitem [{\citenamefont {Vadimov}\ \emph {et~al.}(2019)\citenamefont
  {Vadimov}, \citenamefont {Khaymovich},\ and\ \citenamefont
  {Mel'nikov}}]{1906.02751}%
  \BibitemOpen
  \bibfield  {author} {\bibinfo {author} {\bibfnamefont {V.~L.}\ \bibnamefont
  {Vadimov}}, \bibinfo {author} {\bibfnamefont {I.~M.}\ \bibnamefont
  {Khaymovich}}, \ and\ \bibinfo {author} {\bibfnamefont {A.~S.}\ \bibnamefont
  {Mel'nikov}},\ }\href@noop {} {\enquote {\bibinfo {title} {Higgs modes in
  proximized superconducting systems},}\ } (\bibinfo {year} {2019}),\ \Eprint
  {http://arxiv.org/abs/arXiv:1906.02751} {arXiv:1906.02751} \BibitemShut
  {NoStop}%
\bibitem [{\citenamefont {Kopnin}(2001)}]{KopninBook}%
  \BibitemOpen
  \bibfield  {author} {\bibinfo {author} {\bibfnamefont {N.~B.}\ \bibnamefont
  {Kopnin}},\ }\href@noop {} {\emph {\bibinfo {title} {Theory of Nonequilibrium
  Superconductivity}}}\ (\bibinfo  {publisher} {Oxford University Press},\
  \bibinfo {year} {2001})\BibitemShut {NoStop}%
\bibitem [{\citenamefont {Tsuji}\ and\ \citenamefont
  {Aoki}(2015{\natexlab{b}})}]{PhysRevB.92.064508}%
  \BibitemOpen
  \bibfield  {author} {\bibinfo {author} {\bibfnamefont {N.}~\bibnamefont
  {Tsuji}}\ and\ \bibinfo {author} {\bibfnamefont {H.}~\bibnamefont {Aoki}},\
  }\href {\doibase 10.1103/PhysRevB.92.064508} {\bibfield  {journal} {\bibinfo
  {journal} {Phys. Rev. B}\ }\textbf {\bibinfo {volume} {92}},\ \bibinfo
  {pages} {064508} (\bibinfo {year} {2015}{\natexlab{b}})}\BibitemShut
  {NoStop}%
\bibitem [{\citenamefont {Cea}\ \emph {et~al.}(2016)\citenamefont {Cea},
  \citenamefont {Castellani},\ and\ \citenamefont {Benfatto}}]{Cea2016}%
  \BibitemOpen
  \bibfield  {author} {\bibinfo {author} {\bibfnamefont {T.}~\bibnamefont
  {Cea}}, \bibinfo {author} {\bibfnamefont {C.}~\bibnamefont {Castellani}}, \
  and\ \bibinfo {author} {\bibfnamefont {L.}~\bibnamefont {Benfatto}},\ }\href
  {\doibase 10.1103/PhysRevB.93.180507} {\bibfield  {journal} {\bibinfo
  {journal} {Phys. Rev. B}\ }\textbf {\bibinfo {volume} {93}},\ \bibinfo
  {pages} {180507} (\bibinfo {year} {2016})}\BibitemShut {NoStop}%
\bibitem [{\citenamefont {Silaev}\ \emph
  {et~al.}(2015{\natexlab{b}})\citenamefont {Silaev}, \citenamefont {Virtanen},
  \citenamefont {Heikkil${\rm \ddot a}$},\ and\ \citenamefont
  {Bergeret}}]{Silaev2015a}%
  \BibitemOpen
  \bibfield  {author} {\bibinfo {author} {\bibfnamefont {M.}~\bibnamefont
  {Silaev}}, \bibinfo {author} {\bibfnamefont {P.}~\bibnamefont {Virtanen}},
  \bibinfo {author} {\bibfnamefont {T.~T.}\ \bibnamefont {Heikkil${\rm \ddot
  a}$}}, \ and\ \bibinfo {author} {\bibfnamefont {F.~S.}\ \bibnamefont
  {Bergeret}},\ }\href {\doibase 10.1103/PhysRevB.91.024506} {\bibfield
  {journal} {\bibinfo  {journal} {Phys. Rev. B}\ }\textbf {\bibinfo {volume}
  {91}},\ \bibinfo {pages} {024506} (\bibinfo {year}
  {2015}{\natexlab{b}})}\BibitemShut {NoStop}%
\bibitem [{\citenamefont {Tserkovnyak}\ \emph {et~al.}(2005)\citenamefont
  {Tserkovnyak}, \citenamefont {Brataas}, \citenamefont {Bauer},\ and\
  \citenamefont {Halperin}}]{RevModPhys.77.1375}%
  \BibitemOpen
  \bibfield  {author} {\bibinfo {author} {\bibfnamefont {Y.}~\bibnamefont
  {Tserkovnyak}}, \bibinfo {author} {\bibfnamefont {A.}~\bibnamefont
  {Brataas}}, \bibinfo {author} {\bibfnamefont {G.~E.~W.}\ \bibnamefont
  {Bauer}}, \ and\ \bibinfo {author} {\bibfnamefont {B.~I.}\ \bibnamefont
  {Halperin}},\ }\href {\doibase 10.1103/RevModPhys.77.1375} {\bibfield
  {journal} {\bibinfo  {journal} {Rev. Mod. Phys.}\ }\textbf {\bibinfo {volume}
  {77}},\ \bibinfo {pages} {1375} (\bibinfo {year} {2005})}\BibitemShut
  {NoStop}%
\bibitem [{\citenamefont {Slonczewski}(1996)}]{SLONCZEWSKI1996L1}%
  \BibitemOpen
  \bibfield  {author} {\bibinfo {author} {\bibfnamefont {J.}~\bibnamefont
  {Slonczewski}},\ }\href {\doibase 10.1016/0304-8853(96)00062-5} {\bibfield
  {journal} {\bibinfo  {journal} {J. Magn. Magn. Mater.}\ }\textbf {\bibinfo
  {volume} {159}},\ \bibinfo {pages} {L1} (\bibinfo {year} {1996})}\BibitemShut
  {NoStop}%
\bibitem [{\citenamefont {Brataas}\ \emph {et~al.}(2000)\citenamefont
  {Brataas}, \citenamefont {Nazarov},\ and\ \citenamefont
  {Bauer}}]{PhysRevLett.84.2481}%
  \BibitemOpen
  \bibfield  {author} {\bibinfo {author} {\bibfnamefont {A.}~\bibnamefont
  {Brataas}}, \bibinfo {author} {\bibfnamefont {Y.~V.}\ \bibnamefont
  {Nazarov}}, \ and\ \bibinfo {author} {\bibfnamefont {G.~E.~W.}\ \bibnamefont
  {Bauer}},\ }\href {\doibase 10.1103/PhysRevLett.84.2481} {\bibfield
  {journal} {\bibinfo  {journal} {Phys. Rev. Lett.}\ }\textbf {\bibinfo
  {volume} {84}},\ \bibinfo {pages} {2481} (\bibinfo {year}
  {2000})}\BibitemShut {NoStop}%
\bibitem [{\citenamefont {Waintal}\ \emph {et~al.}(2000)\citenamefont
  {Waintal}, \citenamefont {Myers}, \citenamefont {Brouwer},\ and\
  \citenamefont {Ralph}}]{PhysRevB.62.12317}%
  \BibitemOpen
  \bibfield  {author} {\bibinfo {author} {\bibfnamefont {X.}~\bibnamefont
  {Waintal}}, \bibinfo {author} {\bibfnamefont {E.~B.}\ \bibnamefont {Myers}},
  \bibinfo {author} {\bibfnamefont {P.~W.}\ \bibnamefont {Brouwer}}, \ and\
  \bibinfo {author} {\bibfnamefont {D.~C.}\ \bibnamefont {Ralph}},\ }\href
  {\doibase 10.1103/PhysRevB.62.12317} {\bibfield  {journal} {\bibinfo
  {journal} {Phys. Rev. B}\ }\textbf {\bibinfo {volume} {62}},\ \bibinfo
  {pages} {12317} (\bibinfo {year} {2000})}\BibitemShut {NoStop}%
\bibitem [{\citenamefont {Stiles}\ and\ \citenamefont
  {Zangwill}(2002)}]{PhysRevB.66.014407}%
  \BibitemOpen
  \bibfield  {author} {\bibinfo {author} {\bibfnamefont {M.~D.}\ \bibnamefont
  {Stiles}}\ and\ \bibinfo {author} {\bibfnamefont {A.}~\bibnamefont
  {Zangwill}},\ }\href {\doibase 10.1103/PhysRevB.66.014407} {\bibfield
  {journal} {\bibinfo  {journal} {Phys. Rev. B}\ }\textbf {\bibinfo {volume}
  {66}},\ \bibinfo {pages} {014407} (\bibinfo {year} {2002})}\BibitemShut
  {NoStop}%
\bibitem [{\citenamefont {Bell}\ \emph {et~al.}(2008)\citenamefont {Bell},
  \citenamefont {Milikisyants}, \citenamefont {Huber},\ and\ \citenamefont
  {Aarts}}]{PhysRevLett.100.047002}%
  \BibitemOpen
  \bibfield  {author} {\bibinfo {author} {\bibfnamefont {C.}~\bibnamefont
  {Bell}}, \bibinfo {author} {\bibfnamefont {S.}~\bibnamefont {Milikisyants}},
  \bibinfo {author} {\bibfnamefont {M.}~\bibnamefont {Huber}}, \ and\ \bibinfo
  {author} {\bibfnamefont {J.}~\bibnamefont {Aarts}},\ }\href {\doibase
  10.1103/PhysRevLett.100.047002} {\bibfield  {journal} {\bibinfo  {journal}
  {Phys. Rev. Lett.}\ }\textbf {\bibinfo {volume} {100}},\ \bibinfo {pages}
  {047002} (\bibinfo {year} {2008})}\BibitemShut {NoStop}%
\bibitem [{\citenamefont {Jeon}\ \emph {et~al.}(2018)\citenamefont {Jeon},
  \citenamefont {Ciccarelli}, \citenamefont {Ferguson}, \citenamefont
  {Kurebayashi}, \citenamefont {Cohen}, \citenamefont {Montiel}, \citenamefont
  {Eschrig}, \citenamefont {Robinson},\ and\ \citenamefont
  {Blamire}}]{Jeon2018}%
  \BibitemOpen
  \bibfield  {author} {\bibinfo {author} {\bibfnamefont {K.-R.}\ \bibnamefont
  {Jeon}}, \bibinfo {author} {\bibfnamefont {C.}~\bibnamefont {Ciccarelli}},
  \bibinfo {author} {\bibfnamefont {A.~J.}\ \bibnamefont {Ferguson}}, \bibinfo
  {author} {\bibfnamefont {H.}~\bibnamefont {Kurebayashi}}, \bibinfo {author}
  {\bibfnamefont {L.~F.}\ \bibnamefont {Cohen}}, \bibinfo {author}
  {\bibfnamefont {X.}~\bibnamefont {Montiel}}, \bibinfo {author} {\bibfnamefont
  {M.}~\bibnamefont {Eschrig}}, \bibinfo {author} {\bibfnamefont {J.~W.~A.}\
  \bibnamefont {Robinson}}, \ and\ \bibinfo {author} {\bibfnamefont {M.~G.}\
  \bibnamefont {Blamire}},\ }\href {https://doi.org/10.1038/s41563-018-0058-9}
  {\enquote {\bibinfo {title} {Enhanced spin pumping into superconductors
  provides evidence for superconducting pure spin currents},}\ } (\bibinfo
  {year} {2018})\BibitemShut {NoStop}%
\bibitem [{\citenamefont {Yakushiji}\ \emph {et~al.}(2005)\citenamefont
  {Yakushiji}, \citenamefont {Ernult}, \citenamefont {Imamura}, \citenamefont
  {Yamane}, \citenamefont {Mitani}, \citenamefont {Takanashi}, \citenamefont
  {Takahashi}, \citenamefont {Maekawa},\ and\ \citenamefont
  {Fujimori}}]{Yakushiji2005}%
  \BibitemOpen
  \bibfield  {author} {\bibinfo {author} {\bibfnamefont {K.}~\bibnamefont
  {Yakushiji}}, \bibinfo {author} {\bibfnamefont {F.}~\bibnamefont {Ernult}},
  \bibinfo {author} {\bibfnamefont {H.}~\bibnamefont {Imamura}}, \bibinfo
  {author} {\bibfnamefont {K.}~\bibnamefont {Yamane}}, \bibinfo {author}
  {\bibfnamefont {S.}~\bibnamefont {Mitani}}, \bibinfo {author} {\bibfnamefont
  {K.}~\bibnamefont {Takanashi}}, \bibinfo {author} {\bibfnamefont
  {S.}~\bibnamefont {Takahashi}}, \bibinfo {author} {\bibfnamefont
  {S.}~\bibnamefont {Maekawa}}, \ and\ \bibinfo {author} {\bibfnamefont
  {H.}~\bibnamefont {Fujimori}},\ }\href {https://doi.org/10.1038/nmat1278}
  {\bibfield  {journal} {\bibinfo  {journal} {Nature Materials}\ }\textbf
  {\bibinfo {volume} {4}},\ \bibinfo {pages} {57} (\bibinfo {year}
  {2005})}\BibitemShut {NoStop}%
\bibitem [{\citenamefont {Krause}\ \emph {et~al.}(2007)\citenamefont {Krause},
  \citenamefont {Berbil-Bautista}, \citenamefont {Herzog}, \citenamefont
  {Bode},\ and\ \citenamefont {Wiesendanger}}]{Krause2007}%
  \BibitemOpen
  \bibfield  {author} {\bibinfo {author} {\bibfnamefont {S.}~\bibnamefont
  {Krause}}, \bibinfo {author} {\bibfnamefont {L.}~\bibnamefont
  {Berbil-Bautista}}, \bibinfo {author} {\bibfnamefont {G.}~\bibnamefont
  {Herzog}}, \bibinfo {author} {\bibfnamefont {M.}~\bibnamefont {Bode}}, \ and\
  \bibinfo {author} {\bibfnamefont {R.}~\bibnamefont {Wiesendanger}},\ }\href
  {http://science.sciencemag.org/content/317/5844/1537.abstract} {\bibfield
  {journal} {\bibinfo  {journal} {Science}\ }\textbf {\bibinfo {volume}
  {317}},\ \bibinfo {pages} {1537} (\bibinfo {year} {2007})}\BibitemShut
  {NoStop}%
\bibitem [{\citenamefont {Loth}\ \emph {et~al.}(2010)\citenamefont {Loth},
  \citenamefont {Etzkorn}, \citenamefont {Lutz}, \citenamefont {Eigler},\ and\
  \citenamefont {Heinrich}}]{Loth2010}%
  \BibitemOpen
  \bibfield  {author} {\bibinfo {author} {\bibfnamefont {S.}~\bibnamefont
  {Loth}}, \bibinfo {author} {\bibfnamefont {M.}~\bibnamefont {Etzkorn}},
  \bibinfo {author} {\bibfnamefont {C.~P.}\ \bibnamefont {Lutz}}, \bibinfo
  {author} {\bibfnamefont {D.~M.}\ \bibnamefont {Eigler}}, \ and\ \bibinfo
  {author} {\bibfnamefont {A.~J.}\ \bibnamefont {Heinrich}},\ }\href
  {http://science.sciencemag.org/content/329/5999/1628.abstract} {\bibfield
  {journal} {\bibinfo  {journal} {Science}\ }\textbf {\bibinfo {volume}
  {329}},\ \bibinfo {pages} {1628} (\bibinfo {year} {2010})}\BibitemShut
  {NoStop}%
\bibitem [{\citenamefont {Krause}\ \emph {et~al.}(2016)\citenamefont {Krause},
  \citenamefont {Sonntag}, \citenamefont {Hermenau}, \citenamefont
  {Friedlein},\ and\ \citenamefont {Wiesendanger}}]{PhysRevB.93.064407}%
  \BibitemOpen
  \bibfield  {author} {\bibinfo {author} {\bibfnamefont {S.}~\bibnamefont
  {Krause}}, \bibinfo {author} {\bibfnamefont {A.}~\bibnamefont {Sonntag}},
  \bibinfo {author} {\bibfnamefont {J.}~\bibnamefont {Hermenau}}, \bibinfo
  {author} {\bibfnamefont {J.}~\bibnamefont {Friedlein}}, \ and\ \bibinfo
  {author} {\bibfnamefont {R.}~\bibnamefont {Wiesendanger}},\ }\href {\doibase
  10.1103/PhysRevB.93.064407} {\bibfield  {journal} {\bibinfo  {journal} {Phys.
  Rev. B}\ }\textbf {\bibinfo {volume} {93}},\ \bibinfo {pages} {064407}
  (\bibinfo {year} {2016})}\BibitemShut {NoStop}%
\bibitem [{\citenamefont {Thomas}\ \emph {et~al.}(1996)\citenamefont {Thomas},
  \citenamefont {Lionti}, \citenamefont {Ballou}, \citenamefont {Gatteschi},
  \citenamefont {Sessoli},\ and\ \citenamefont {Barbara}}]{Thomas1996}%
  \BibitemOpen
  \bibfield  {author} {\bibinfo {author} {\bibfnamefont {L.}~\bibnamefont
  {Thomas}}, \bibinfo {author} {\bibfnamefont {F.}~\bibnamefont {Lionti}},
  \bibinfo {author} {\bibfnamefont {R.}~\bibnamefont {Ballou}}, \bibinfo
  {author} {\bibfnamefont {D.}~\bibnamefont {Gatteschi}}, \bibinfo {author}
  {\bibfnamefont {R.}~\bibnamefont {Sessoli}}, \ and\ \bibinfo {author}
  {\bibfnamefont {B.}~\bibnamefont {Barbara}},\ }\href
  {https://doi.org/10.1038/383145a0} {\bibfield  {journal} {\bibinfo  {journal}
  {Nature}\ }\textbf {\bibinfo {volume} {383}},\ \bibinfo {pages} {145}
  (\bibinfo {year} {1996})}\BibitemShut {NoStop}%
\bibitem [{\citenamefont {Bobkova}\ and\ \citenamefont
  {Bobkov}(2017)}]{PhysRevB.96.104515}%
  \BibitemOpen
  \bibfield  {author} {\bibinfo {author} {\bibfnamefont {I.~V.}\ \bibnamefont
  {Bobkova}}\ and\ \bibinfo {author} {\bibfnamefont {A.~M.}\ \bibnamefont
  {Bobkov}},\ }\href {\doibase 10.1103/PhysRevB.96.104515} {\bibfield
  {journal} {\bibinfo  {journal} {Phys. Rev. B}\ }\textbf {\bibinfo {volume}
  {96}},\ \bibinfo {pages} {104515} (\bibinfo {year} {2017})}\BibitemShut
  {NoStop}%
\bibitem [{\citenamefont {Silaev}(2019{\natexlab{b}})}]{PhysRevB.99.224511}%
  \BibitemOpen
  \bibfield  {author} {\bibinfo {author} {\bibfnamefont {M.}~\bibnamefont
  {Silaev}},\ }\href {\doibase 10.1103/PhysRevB.99.224511} {\bibfield
  {journal} {\bibinfo  {journal} {Phys. Rev. B}\ }\textbf {\bibinfo {volume}
  {99}},\ \bibinfo {pages} {224511} (\bibinfo {year}
  {2019}{\natexlab{b}})}\BibitemShut {NoStop}%
\bibitem [{\citenamefont {Murotani}\ and\ \citenamefont
  {Shimano}(2019)}]{PhysRevB.99.224510}%
  \BibitemOpen
  \bibfield  {author} {\bibinfo {author} {\bibfnamefont {Y.}~\bibnamefont
  {Murotani}}\ and\ \bibinfo {author} {\bibfnamefont {R.}~\bibnamefont
  {Shimano}},\ }\href {\doibase 10.1103/PhysRevB.99.224510} {\bibfield
  {journal} {\bibinfo  {journal} {Phys. Rev. B}\ }\textbf {\bibinfo {volume}
  {99}},\ \bibinfo {pages} {224510} (\bibinfo {year} {2019})}\BibitemShut
  {NoStop}%
\bibitem [{\citenamefont {Heikkil{\"a}}\ \emph {et~al.}(2018)\citenamefont
  {Heikkil{\"a}}, \citenamefont {Ojaj{\"a}rvi}, \citenamefont {Maasilta},
  \citenamefont {Strambini}, \citenamefont {Giazotto},\ and\ \citenamefont
  {Bergeret}}]{heikkila2018thermoelectric}%
  \BibitemOpen
  \bibfield  {author} {\bibinfo {author} {\bibfnamefont {T.~T.}\ \bibnamefont
  {Heikkil{\"a}}}, \bibinfo {author} {\bibfnamefont {R.}~\bibnamefont
  {Ojaj{\"a}rvi}}, \bibinfo {author} {\bibfnamefont {I.~J.}\ \bibnamefont
  {Maasilta}}, \bibinfo {author} {\bibfnamefont {E.}~\bibnamefont {Strambini}},
  \bibinfo {author} {\bibfnamefont {F.}~\bibnamefont {Giazotto}}, \ and\
  \bibinfo {author} {\bibfnamefont {F.~S.}\ \bibnamefont {Bergeret}},\ }\href
  {\doibase 10.1063/1.5037405} {\bibfield  {journal} {\bibinfo  {journal}
  {Phys. Rev. Appl.}\ }\textbf {\bibinfo {volume} {10}},\ \bibinfo {pages}
  {034053} (\bibinfo {year} {2018})}\BibitemShut {NoStop}%
\end{thebibliography}%

\end{document}